\documentclass[useAMS,usenatbib]{mn2e}
\usepackage{graphicx,rotate,url,mathptmx,lscape,times,amsmath,amssymb}
\def\gtsim{\mathrel{\hbox{\rlap{\hbox{\lower4pt\hbox{$\sim$}}}\hbox{$>$}}}}
\def\lesssim{\mathrel{\hbox{\rlap{\hbox{\lower4pt\hbox{$\sim$}}}\hbox{$<$}}}}

\def\Msun{M$_{\odot}$}
\def\ga{{\rm\thinspace gauss}}

\def\km{{\rm\thinspace km}}

\def\Msun{\hbox{$\rm\thinspace M_{\odot}$}}

\def\ps{{\rm\thinspace s^{-1}}}

\def\kmps{\hbox{$\km\ps\,$}}

\def\J{\hbox{$J$}}
\def\Ks{\hbox{$Ks$}}
\def\ktot{\hbox{$K_{tot}$}}
\def\H{\hbox{$H$}}
\def\MRCttf{\hbox{MRC\,1324\,--\,262}}
\def\MRCofos{\hbox{MRC\,0406\,--\,244}}
\def\MRCtoof{\hbox{MRC\,2104\,--\,242}}
\def\MRCtotn{\hbox{MRC\,2139\,--\,292}}
\def\USS{\hbox{USS\,1425\,--\,148}}
\def\MG{\hbox{MG\,2308\,$+$\,0336}}

\def\green{\hbox{\it JHK}}
\def\black{\hbox{\it ALL-JHK}}

\usepackage{color}
\def\h0{\hbox{{\rm H}$^0$}}
\DeclareMathAlphabet{\vib}{OML}{cmm}{m}{it}

\begin{document}

\title[Galaxy overdensities at z$\sim2.4$]{Galaxy protocluster candidates around $z\sim2.4$ radio galaxies.}

\author[N. A. Hatch et al.]
       {\parbox[]{6.0in}
       {N.\,A.\,Hatch$^{1,2}$\thanks{E-mail: nina.hatch@nottingham.ac.uk},~C.\,De\,Breuck$^{3}$,~A.\,Galametz$^{3}$,~G.\,K.\,Miley$^{1}$, R.\,A.\,Overzier$^4$, H.\,J.\,A.\,R\"ottgering$^{1}$, M.\,Doherty$^5$,  T.\,Kodama$^{6}$, J.\,D.\,Kurk$^7$, N.\,Seymour$^8$, B.\,P.\,Venemans$^{3}$, J.\,Vernet$^{3}$, and A.\,W.\,Zirm.$^9$\\
        \footnotesize
        $^1$Leiden Observatory, University of Leiden, P.B. 9513, Leiden 2300 RA, The Netherlands\\
        $^2$School of Physics and Astronomy, University of Nottingham, University Park, Nottingham NG7 2RD\\
        $^3$European Southern Observatory, Karl-Schwarzschild-Str. 2, D-85748 Garching, Germany\\
        $^4$Max-Planck-Institut f{\"u}r Astrophysik, Karl-Schwarzschild Strasse 1, D-85741 Garching, Germany\\
        $^5$European Southern Observatory, ESO Santiago, Alonsode Cordova 3107, Vitacura, Santiago, Chile \\
        $^6$National Astronomical Observatory of Japan, Mitaka, Tokyo 181-8588, Japan\\
        $^7$Max-Planck-Institut f{\"u}r Extraterrestrische Physik, Giessenbachstrasse, D-85741 Garching, Germany\\
        $^8$Mullard Space Science Laboratory, UCL, Holmbury St Mary, Dorking, Surrey, RH5 6NT, UK\\
        $^9$Dark Cosmology Centre, Niels Bohr Institute, University of Copenhagen, Juliane Maries Vej 30, DK-2100 Copenhagen, Denmark.
        }}

\date{Accepted 
      Received }

\pubyear{}

\maketitle

\label{firstpage}
\begin{abstract}
We study the environments of 6 radio galaxies at $2.2<z<2.6$ using wide-field near-infrared images. We use colour cuts to identify galaxies in this redshift range, and find that three of the radio galaxies are surrounded by significant surface overdensities of such galaxies. The excess galaxies that comprise these overdensities are strongly clustered, suggesting they are physically associated. The colour distribution of the galaxies responsible for the overdensity are consistent with those of galaxies that lie within a narrow redshift range at $z\sim2.4$. Thus the excess galaxies are consistent with being companions of the radio galaxies. The overdensities have estimated masses in excess of $10^{14}$\Msun, and are dense enough to collapse into virizalised structures by the present day: these structures may evolve into groups or clusters of galaxies. 

A flux-limited sample of protocluster galaxies with $K<20.6$\,mag is derived by statistically subtracting the fore- and background galaxies.  The colour distribution of the protocluster galaxies is bimodal, consisting of a dominant blue sequence, comprising 77$\pm10$\% of the galaxies, and a poorly populated red sequence. The blue protocluster galaxies have similar colours to local star-forming irregular galaxies ($U-V _{\rm AB}\sim0.6$), suggesting most protocluster galaxies are still forming stars at the observed epoch. The blue colours and lack of a dominant protocluster red sequence implies that these cluster galaxies form the bulk of their stars at $z\lesssim3$. 

\noindent 

\end{abstract}

\section{Introduction}
In the local Universe galaxy clusters are dominated by red early-type galaxies \citep{Dressler1980} that sit on a tight linear sequence in colour-magnitude space known as the red sequence \citep{Visvanathan1977}. These galaxies contain passively evolving stellar populations, and have little on-going star formation. The intrinsic scatter of galaxies about the red sequence is small \citep[e.g.,][]{Bower1992,Stanford1998}, and both the scatter and slope of the sequence does not evolve with redshift \citep{Blakeslee2003,Gladders1998}. These results suggest that the stars in cluster elliptical galaxies formed in a brief but vigorous epoch of star formation.  Current estimates place this epoch of intense star formation at $z=2-3$, which we can thus assume to be an important time in the formation of clusters  \citep{vanDokkum2007,Kurk2009,Papovich2010,Tanaka2010}.  

A protocluster is a large-scale overdense region in the early Universe, whose properties are consistent with it being a progenitor of a local galaxy cluster. Such objects are observed at a time when their member galaxies are not yet virialized within single cluster-sized dark matter halos. To qualify as a protocluster, a given overdense region must have a mass comparable to that of local clusters of galaxies ($M\gtrsim10^{14}$\Msun), and be dense enough to collapse before $z=0$.

An efficient technique for finding the progenitors of clusters involves targeting high redshift radio galaxies (HzRGs). Galaxy formation is expected to be more efficient in high density environments than in the field, due to the abundance of surrounding gas \citep{Gunn1972}.  Therefore one may expect HzRGs, which are among the most massive galaxies in the early Universe \citep{Rocca-Volmerange2004,Seymour2007}, to pinpoint the location of the highest density regions, such as protoclusters.  Searches for galaxy overdensities near these objects have resulted in detections of large-scale structures and protoclusters up to $z=5.2$ \citep[e.g.,][]{LeFevre1996,Pentericci2000,Kurk2000,Kurk2004,Best2003,Overzier2006,Venemans2007,Overzier2008}.

We have performed an infrared survey of the fields of 8 powerful radio galaxies with redshifts in the range $1.7\le z \le 2.6$ using the wide-field near-infrared imager HAWK-I on the VLT. The aim of the survey is to identify protoclusters and study the member galaxies and their evolution over this redshift range.  The HAWK-I data on the two lowest redshift targets MRC\,$1017-220$ and MRC\,$0156-252$ are presented in \citet{Galametz2010}.
 
In this paper we present HAWK-I data for the 6 highest redshift targets, ranging from $2.2 \le z \le 2.6$. In Section \ref{obs_dr} we describe the observations and data reduction. Sections \ref{identify} -- \ref{stat} lay out the evidence that 3 out of 6 of the radio galaxies lie within galaxy protoclusters. In Section \ref{discussion} we discuss the properties of the overdensities, and the colours of the galaxies within the 3 protocluster candidates. 

Fluxes are calibrated using the Vega magnitude scale unless noted otherwise. A flat $\Lambda$CDM cosmology is assumed throughout, with $\Omega_{\rm M}=0.3$, $\Omega_{\Lambda}=0.7$ and H$_0$=70\kmps Mpc$^{-1}$. The distance scale at $z=2.4$ is 8.139\,kpc/\arcsec, or $\sim$1.7\,co-moving Mpc/\arcmin. Distances are given in co-moving units, unless stated otherwise.

\section{Observations and data reduction}
\label{obs_dr}
\subsection{Target selection and observations}
Six HzRGs with redshifts between 2.28 and 2.55 were selected from the compendium of \citet{MileydeBreuck2008}. The targets were chosen based on their distributions in right ascension, declination, and redshift, and their bright radio luminosity  (L$_{500 \rm{MHz}}>10^{28.5}$W\,Hz$^{-1}$), and not on any prior information about surrounding  galaxy overdensities. Co-ordinates of the selected targets and the control field are given in Table~\ref{tab:obs}. 

The 6 HzRG fields and a blank control field were observed in service mode using the High Acuity Wide field K-band Imager (HAWK-I; \citealt{Kissler-Patig2008}) on the ESO Very Large Telescope (VLT) UT4-YEPUN telescope in Paranal, Chile during the period April--September 2008.   
HAWK-I is a near-infrared camera comprising of four Hawaii-2 2048x2048 pixel detectors separated by a gap of $\sim$15\arcsec. The camera spans $7.5\times7.5$\,arcmin with a pixel scale of 0.106 arcsec per pixel. 

The telescope pointing was optimised so that the HzRG were placed close to the centre of the HAWK-I field of view, but at least 1\,arcmin away from the chip gaps, and bright stars did not fall within the field of view. Each target was observed through the \J, \H\ and \Ks\ filters and total exposure times for all targets are provided in Table.\,\ref{tab:obs}.  The telescope dithered every 2\,mins which resulted in a $\sim$1\,arcmin-wide cross-shaped region on the final mosaic where the image depth is shallower. 

During the observing period the anti-reflection coating of the Dewar window of HAWK-I was damaged resulting in small cross-shaped patterns on the exposures. These patterns, caused by the spider of the secondary mirror, rotated as the telescope tracked the targets as HAWK-I is situated on the Nasmyth focus. These patterns were not adequately removed by the background subtraction step unless the telescope had not moved far between successive exposures. Thus the integration time between dithers was reduced from 2\,mins to 1\,min for all images obtained after May 2008.

\begin{table*}
  \centering
  \caption{Summary of the target fields and the HAWK-I observations. 500\,MHz radio luminosities, given in column 3, are from \citet{MileydeBreuck2008}. \label{tab:obs}}
 \begin{tabular}{lcccccrcc}
  \hfill
 Radio galaxy &Radio galaxy &Radio luminosity&Centre of field &  Redshift   &Filter & Exposure time& 5$\sigma$  depth & Seeing \\
   &coordinates (J2000)& Log(500\,MHz)&(J2000) & & &(hours)& (Vega) & (FWHM)\\  \hline
\MRCttf&13:26:54.6  --26:31:42&28.46&13:26:51  --26:30:55&2.28&\J&3.42&24.9&0.7\\
&&&&&\H&0.88&23.6&0.7\\
&&&&&\Ks&0.42&22.0&0.42\\
\USS&14:28:41.7 --15:02:28.4&28.66&14:28:53  --15:01:43&2.35&\J&3.15&25.0&0.65\\
&&&&&\H&0.72&23.0&0.6\\
&&&&&\Ks&0.56&22.0&0.6\\
\MRCofos&04:08:51.5  --24:18:16.7&29.03&04:08:43  --24:17:35&2.43&\J&3.38&24.8&0.4\\
&&&&&\H&0.84&23.5&0.7\\
&&&&&\Ks&0.53&22.8&0.74\\
\MG&23:08:25.1 +03:37:03.9&28.51&23:08:28 +03:36:15&2.46&\J&3.40&25.5&0.55\\
&&&&&\H&0.88&23.4&0.55\\
&&&&&\Ks&0.56&22.4&0.5\\
\MRCtoof&21:06:58.3  --24:05:09.0&28.84&21:06:55 --24:04:31&2.49&\J&3.38&25.3&0.65\\
&&&&&\H&0.67&23.8&0.75\\
&&&&&\Ks&1.53&22.6&0.6\\
\MRCtotn&21:42:16.6 --28:58:38.2&28.74&21:42:10  --28:59:32&2.55&\J&3.58&25.3&0.45\\
&&&&&\H&1.70&24.0&0.55\\
&&&&&\Ks&0.53&22.7&0.6\\
Control field&--&--&01:28:04  --26:26:08&--&\J&3.18&25.2&0.7\\
&&&&&\H&0.84&23.5&0.65\\
&&&&&\Ks&0.53&22.7&0.7\\
\hline
\end{tabular}
 
\end{table*}
\subsection{Data reduction}

\label{dr}
The data were reduced using the ESO/MVM \citep{Vandame2004} data reduction pipeline optimised for the reduction of our HAWK-I data. The usual near-infrared reduction steps were taken, including dark subtraction, flat-field removal, harmonising the gain of the four detectors, removal of fringing, sky-subtraction, creation of bad pixel and weight maps, calculating the relative astrometry between the chips and the absolute astrometry. 

USNO-B1 catalogues \citep{Monet2003} were used to calculate the first-guess astrometric solutions for the \Ks\ images. A catalogue was then compiled from the objects detected in the \Ks\ images and used to calculate the astrometric solution of the $J$ and $H$ images. The accuracy of the relative astrometry between the 3 images of each target is typically within 0.1\,pixel, but the absolute astrometry is limited by the USNO-B1 catalogue which has an accuracy of  0.2\,arcsec, equivalent to 2 HAWK-I pixels.

After the sky background was removed from each exposure by MVM in a two step process, low level large-scale variations were still seen across each quadrant and most notably between the quadrants of the reduced images. Therefore the sky background of the final images was measured and subtracted using a local background estimator with SExtractor ({\sc BACKPHOTO\_TYPE LOCAL}).

The data were flux calibrated with 2MASS catalogues, using $13-15.5$\,mag stars within the target fields. The calibration was checked using $13-14$\,mag stars within standard star fields taken within 2\,hours of the observations. Typically the zero-points determined from both methods agreed within the uncertainties ($\sim0.05$\,mag). 

The end products of this reduction process are a science image containing the reduced data of the target, and an effective exposure-time image, which is an exposure-time map that has been normalised to account for differences in sensitivity between the four detectors of HAWK-I and the chip gaps. 

The 3 colour images of a target were convolved to match the lowest resolution image using the {\sc iraf} package {\sc psfmatch}. The convolution kernels were optimised so that the stellar growth curves (created by median combining at least 20 bright and unsaturated stars in each image) converged to within 3\% at and beyond a 1$\arcsec$ radius. The 5$\sigma$ image depths given in Table\,\ref{tab:obs} were measured by placing 2$\arcsec$ diameter apertures at multiple random positions. 

\subsection{Masking of cross-talk, bright objects, and regions with low exposure time}
Each of the four HAWK-I detectors contain 32 amplifiers. Cross-talk between the amplifiers produce a series of artefacts arranged horizontally with respect to each star in the field at regular 64 pixel intervals \citep{finger2008}. These artefacts appear crater-like and although are produced for every object in the field, they are only detectable above the noise if the star is brighter than approximately $J=14.5$\,mag (although this is strongly dependent on the seeing). The artefacts were most pronounced in the deep \J\ images.  $2\arcsec\times1\arcsec$ rectangular regions were masked at 64 pixels intervals from each star brighter than $J\simeq14.5$ across the entire detector quadrant. Cross-talk was greatly reduced in programs observed after ESO semester 82. 

Bright stars and nearby galaxies were also masked because they cover a significant amount of area in some fields and therefore reduce the total area where faint galaxies could be detected. All objects brighter than $Ks=16$\,mag were masked out.  Finally all pixels with less than 30 percent of the maximum effective exposure time were masked as these regions are incomplete even at bright magnitudes and could lead to spurious detections.

\subsection{Source detection}
Sources were detected using SExtractor \citep{Bertin1996} in double image mode using weighted \Ks\ images as detection images. The weighted images, created by multiplying the reduced \Ks\ images by the square root of their associated effective exposure-time maps, take into account the varying background noise level. Sources were defined as groups of 5 pixels that are 1.25$\sigma$ above the local background noise. The data were filtered, before object detection, with Gaussian filters of width and size that matched the seeing. 

The colour of objects were measured from the PSF-matched images using apertures of approximately 2\,arcsec diameter.\footnote{The exception is the \MRCtotn\ field, for which 2.2\arcsec\ apertures were used, since the stellar growth curves of the 3 images only converge to less than 3\% beyond this diameter.} Total  $Ks$ fluxes were measured from the high-resolution $Ks$ images using Kron apertures (FLUX\_AUTO parameter of SExtractor). 

Photometry in Kron apertures can underestimate the total flux, especially if the source is faint. Therefore aperture corrections were applied to convert the $Ks$ Kron fluxes to total \Ks\ fluxes (\ktot). Correction factors were determined from stellar growth curves of bright unsaturated stars in the \Ks\ images: a circular aperture, of the same total area as the Kron aperture of each object, was used to measure the amount of light missed by the Kron aperture. The amount of aperture correction depends on the PSF of the image, and on the source size and brightness, but the source flux was typically increases by this correction by a few percent, up to 20\% for the smallest objects. 
 
Uncertainties on the fluxes in each band were estimated through $\sigma^2=5\sigma^2_{\rm bkgrd} + \sigma^2_{\rm Poisson}$, where $\sigma_{\rm bkgrd}$ is the image background noise and $\sigma_{\rm Poisson}$ is the Poisson uncertainty in the number of detected electrons (flux in ADU$\times$effective instrument gain) measured by SExtractor.

\subsection{Image completeness}
\label{comp}

The point-source and galaxy completion of the \Ks\ HAWK-I data was calculated by simulating 3000 point sources or galaxies on each image (in batches of 1000) using the {\sc iraf} packages {\sc gallist} and {\sc mkobject}. The simulated galaxies were set to have a 2.5 pixel half-light radius, as observed for bright galaxies in the images, and we assume the elliptical-to-spiral fraction is 0.4, and maximum elliptical axial ratio is 0.3. 

The image completion is uneven due to the varying effective exposure time, so each simulated source was assigned an effective exposure time calculated from the weights map. The sources were then detected with SExtractor, grouped into 4 bins of effective exposure time, and the completion level calculated for each bin. Regions with low effective exposure time have lower completion levels than regions with high effective exposure times. The average completion for the entire image was created by averaging the completion of the 4 exposure-time categories, weighted by the fraction of image area that lies within each exposure time category. 

\begin{figure}
 \begin{center}
 \includegraphics[width=1\columnwidth]{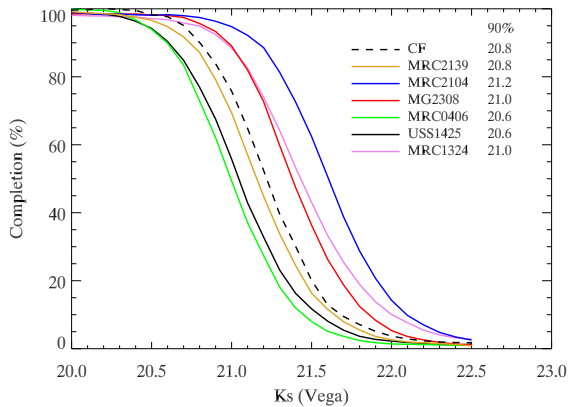}
\caption{The average completeness of extended sources in all 7 observed \Ks\ fields.   The images are complete to $\sim0.6$\,mag fainter for point sources than extended sources.  The entire image is complete to at least the 90\% limiting magnitude, including regions with low effective exposure time. Below this limit the galaxy completeness is not uniform across the images. 90\% galaxy completeness limits are given in the legend. \label{fig:complete_gal_K}}
\end{center}
\end{figure}

The average galaxy completion of each image is shown in Fig.\,\ref{fig:complete_gal_K}. The 90\% completeness is the limit to which the entire image is complete to at least 90\%, including regions with low effective exposure time. Below this limit the galaxy completeness is not uniform across the images because of the varying effective exposure time. The images are complete to $\sim0.6$\,mag fainter for point sources.

\label{depth}

\subsection{Stellar classification}
At bright magnitudes ($K<19$ mag), SExtractor can adequately select unresolved objects based on their morphology, but this selection method is not adequate at faint magnitudes. Therefore objects were classified as stars based on their near-infrared colours. The near-infrared colours of bright stars were examined in the colour-magnitude diagrams of the 7 observed fields, and the following colour-selection criteria were designed to separate stars from galaxies. Stars were defined as objects which obey all of the following conditions:
\begin{eqnarray}
J-H< 0.55~~\cap~~H-K< 0.85~~\cap~~J-K<1.15~. 
\end{eqnarray}
These criteria were tested using the Pickles stellar library. All stars are selected by this criteria, except for very red stars such as  L- or early T-type dwarfs, but the density of such stars is very low. 

We identify stars in both the control fields and the HzRG fields using the same colour criteria so the results presented in this work are not sensitive to moderate changes in these criteria.  Changing these colour cuts by up to 0.1\,mag does not affect our conclusions.

\section{Identifying overdensities in the radio galaxy fields}
\label{identify}

\subsection{Galaxy number counts}
\begin{figure}
 \begin{center}
 \includegraphics[width=1\columnwidth]{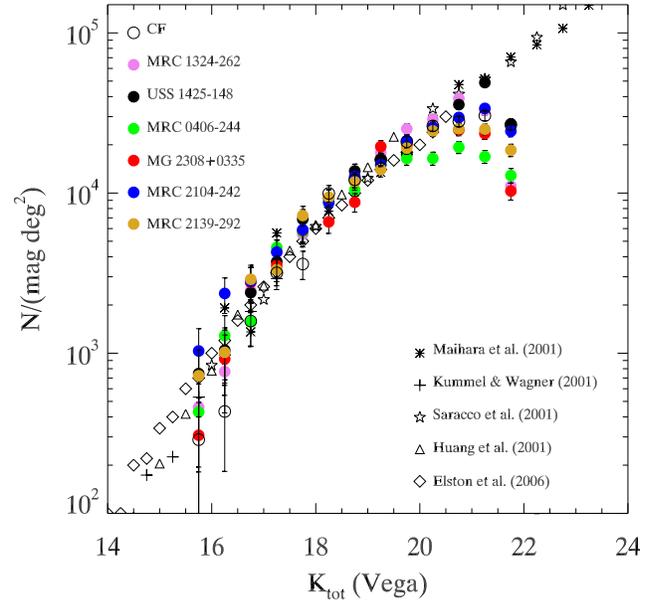}
 \caption{Galaxy number counts in $K$ for the 6 radio galaxy fields compared to the control field (CF) and values taken from the literature. No completeness correction was applied to the radio galaxy fields or CF. There is good agreement between the radio galaxy fields and data from the literature, in particular between 17.5 and 20.5 mag.  The radio galaxy fields do not contain a large excess of galaxies. \label{fig:nc}}
\end{center}
\end{figure}
Differential $K$-band galaxy number counts were measured from the HzRG fields and the control field (CF)  and are compared to galaxy counts from the literature in Fig.\,\ref{fig:nc}. The galaxy number counts were not corrected for completeness, and no attempt was made to correct for the difference in the filter bandpasses. 

The observed number counts are fully consistent with literature counts up to the completion limit of the images: there is no large galaxy excess in any HzRG field compared to the control field or the literature. Any protocluster structure associated with the HzRGs may only contribute a small fraction of the galaxies in the observed field.  

The wide-field HAWK-I images contain a large contribution of foreground and background objects which dominate over any structure associated with the radio galaxy.  Additionally, cosmic variance of these galaxies may hide a protocluster, or make a region appear overdense through chance superposition. It is therefore necessary to remove as many interloping galaxies as possible before searching for structures around the HzRGs. 

\subsection{Selecting galaxies at $2.2<z<2.7$}
\label{sec:colours_z24}
\begin{figure}
 \begin{center}
 \includegraphics[width=0.9\columnwidth]{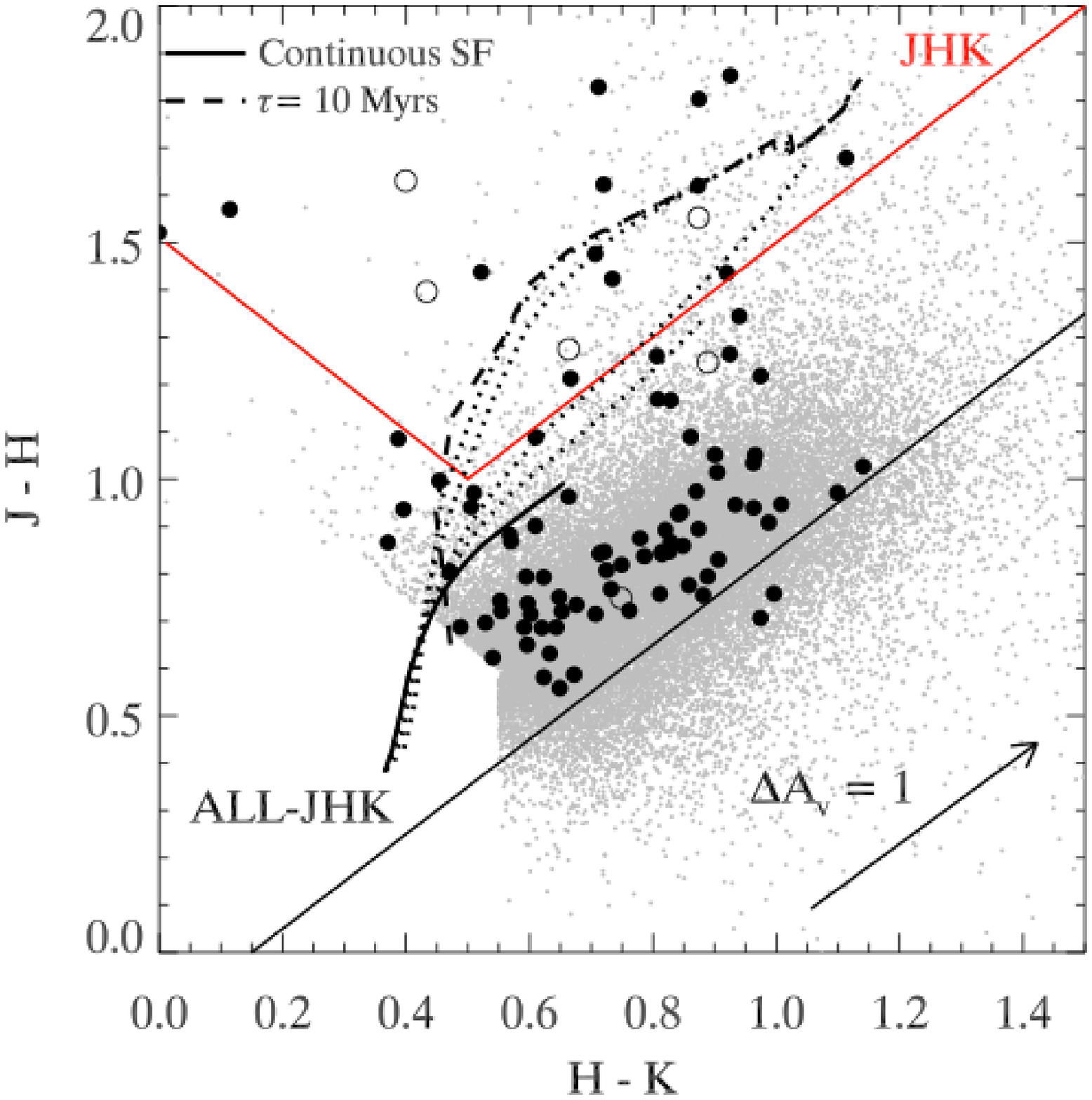}
 \caption{Near-IR colours of galaxies in the 0.8\,deg$^2$ UDS (grey circles), UDS galaxies with $2.2<z_{phot}<2.7$ (black solid circles), and the 6 observed HzRGs (open circles). Red lines mark the \green\ criteria designed to select galaxies at the same redshift as the HzRGs. 98\% of UDS galaxies with $2.2<z_{phot}<2.7$ satisfy the \black\ criterion (solid black line). \citet{BC03} models of galaxies at $z=2.4$ are overplotted. Each track corresponds to galaxies with ages evolving from $0.01$ to $2.6$~Gyr ($J-H$ reddens with ages). The 6 tracks have different star formation histories: continuous star formation (solid line), and exponentially declining star formation (dotted lines) with  $\tau=1$,$0.5$,$0.1$ and $0.05$~Gyr (from bottom to top) and $\tau=0.01$~Gyr (dashed line). The reddening vector is shown by the black arrow. \label{fig:JH_HK_track}}
\end{center}
\end{figure}

To identify galaxy overdensities near the HzRGs we attempt to isolate galaxies at $2.2<z<2.7$.  For galaxies in this redshift range the Balmer and 4000\AA\ break lies between the $J$ and $H$ filter bandpasses.

\subsubsection{Colours of galaxies with photometric redshifts in the range $2.2<z_{phot}<2.7$}
\label{UDS_phot_section}
Fig.\,\ref{fig:JH_HK_track} plots the near-IR colours of all galaxies within the 0.8\,deg$^2$ $K-$selected Ultra Deep Survey (UDS; \citealt{Lawrence2007}). This field is covered by deep $B$, $V$, $R$, $i^\prime$, z$^{\prime}$, $J$, $H$, $K$,  {\it Spitzer} $3.6\mu$m and $4.5\mu$m data. An updated version of the $K$-selected catalogue of \citet{Williams2009a} is used, in which the data from UKIDSS Data release 1 is supplemented with $H$-band data from the UKIDSS Data release 3 (see \citealt{Williams2009b} for details). In addition to the photometric catalogues we use the photometric redshift catalogues compiled in \citet{Williams2009b}. We refer the reader to \citet{Williams2009a,Williams2009b} for further details on both catalogues.

A sample of galaxies were selected from the UDS which have photometric redshifts in the range $2.2<z<2.7$.  These galaxies have similar redshifts as the HzRGs, and are likely to have similar colours to the companions of the radio galaxies. To obtain the correct galaxy colours, it is essential that the Balmer/4000\AA\ break of the UDS galaxies falls between the $J$ and $H$ bands, just like the HzRGs. Therefore we only select galaxies whose redshift fits have $\log\,\chi^2<2.9$, and a narrow redshift probability distribution, such that the 3$\sigma$ confidence intervals of the redshift estimate lie within the redshift range $2.2<z<2.7$. This resulted in a sample of 89 galaxies with $K_{tot}<20.6$\,mag. These galaxies are plotted in Fig.\,\ref{fig:JH_HK_track} as filled black circles. Hereafter this sample is referred to as $z_{\rm phot}$-UDS.

\subsubsection{Colour criteria to select $2.2<z<2.7$ galaxies}
\citet{Kajisawa2006} and \citet{Kodama2007} devised a 2-colour selection criterion,
\begin{equation}
J-H> H-K +0.5 ~\cap~ ~J-K>1.5 ~~~~~~~~~~~~~(\green).
\nonumber
\label{green}
\end{equation}
which selects galaxies whose Balmer or 4000\AA\ breaks are bracketed by the $J$ and $H$ passbands, so it is well suited to our purpose of selecting galaxies at the same redshift as the $z\sim2.4$ HzRGs.  Galaxies selected by this criterion are referred to as  \green\ galaxies following the convention of \citet{Kajisawa2006}. 

Approximately half (53\%) of UDS galaxies who satisfy the \green\ criteria have photometric redshifts in the range $2.2<z_{phot}<2.7$. This selection technique has also been spectroscopically tested by \citet{Doherty2010}, who found that 10 out of 18 \green\ galaxies lie in the redshift range $2.2<z<2.7$. Therefore approximately half of the \green\ galaxies in the HzRG fields are likely to lie near the redshift of the HzRG.

This \green\ criterion is biased toward selecting galaxies with low star formation rates, and cannot detect $z\sim2.4$ galaxies with continuous star formation histories (see Fig.\,\ref{fig:JH_HK_track}). 
Over 85\% of the UDS galaxies with photometric redshifts in the range $2.2<z_{phot}<2.7$ do not meet the \green\ criteria, so many protocluster galaxy candidates may be missed. We therefore define a second colour criterion
\begin{equation}
~~~~~~~J-H > H-K-0.15,~~~~~~~~~~~~~~~~~(\black),
\nonumber
\label{red}
\end{equation}
which is marked by the solid black line in Fig.\,\ref{fig:JH_HK_track}.  98\% of UDS galaxies with photometric redshifts in the range $2.2<z_{phot}<2.7$ fall within this colour criterion. However only a minority of this sample will lie at the same redshift as the HzRGs, since only 4\% of the UDS galaxies that satisfy this criterion have photometric redshifts in the range $2.2<z_{phot}<2.7$.

Luminous radio galaxies are among the brightest galaxies at all redshifts \citep{Rocca-Volmerange2004} so accompanying protocluster galaxies are likely to be fainter than the HzRGs. The brightest HzRG has $K_{tot}=17.7$\,mag, so we assume that all galaxies with $K_{tot}<17.5$\,mag are interlopers. Only the \MRCtoof\ field contains an excess of galaxies with $K_{tot}<17.5$\,mag (see Fig.\,\ref{fig:nc}). Visual inspection of these galaxies reveal that they have large angular sizes and blue near-infrared colours, so they are likely to be low redshift interlopers. For the rest of this work galaxies with $K_{tot}<17.5$\,mag are removed from the catalogues of the HzRG fields and control fields.

\subsection{Comparison control fields}
\label{comparision_fields}
\begin{figure}
\begin{center}
\includegraphics[width=0.475\columnwidth]{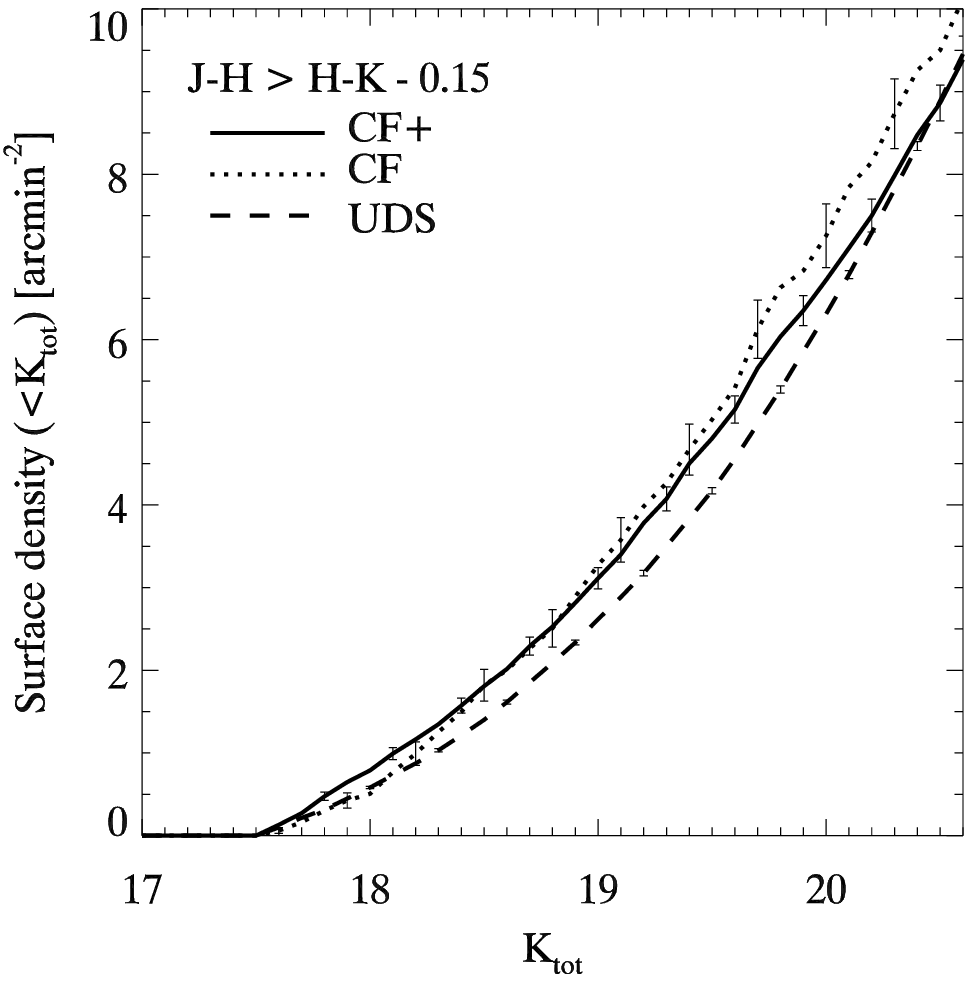}
\includegraphics[width=0.475\columnwidth]{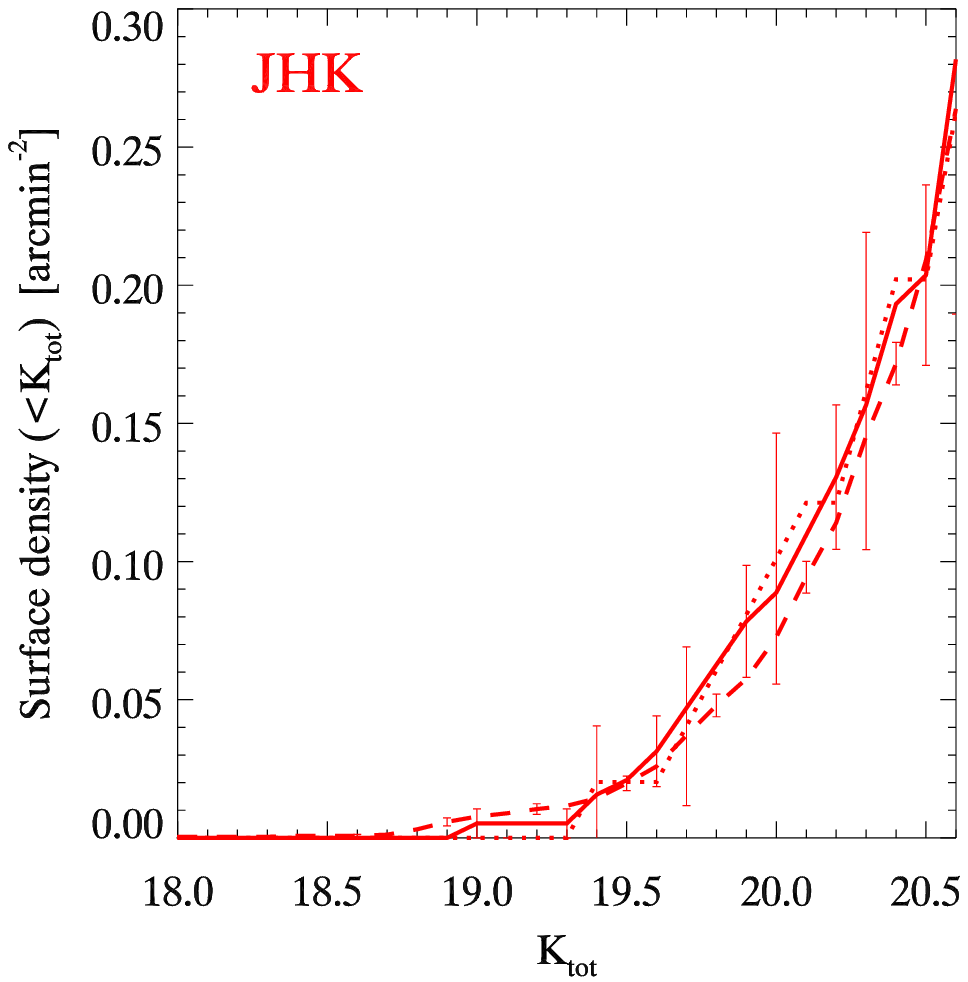}	
\caption{Comparison of the 3 control fields: the solid line is the  230\,arcmin$^{2}$ control field (CF+) consisting of the observed control field and the outer regions of the radio galaxy fields; the dotted line is the 56\,arcmin$^{2}$ observed control field (CF); and dashed line is 0.8\,deg$^2$ UDS.  Error bars come from Poisson statistics. There is good agreement between all 3 control fields for both colour-selected galaxy populations. \label{fig:comparison}}
\end{center}
\end{figure}
Reference control fields are required to identify overdensities in the radio galaxy fields. A control field (CF) was observed with the same instrumental set-up as the radio galaxy fields, allowing for a direct comparison without observational bias. This control field is a single HAWK-I field-of-view  (56\,arcmin$^{2}$) so is sensitive to cosmic variance. Therefore a larger control field was compiled by supplementing the CF with the outer regions of the 6 radio galaxy fields i.e.~the regions beyond 3\,arcmin ($\sim5$\,Mpc co-moving at $z=2.4$) of the HzRGs. This combined control field results in a total area of 230\,arcmin$^{2}$, hereafter referred to as CF+.

The cumulative surface density within the control field (CF), CF+ ,  and the 0.8\,deg$^{2}$ UDS are compared in Fig.\,\ref{fig:comparison}.  The surface density of galaxies in the CF+ is generally lower than or at the same level as the CF for all magnitudes, so the CF+ field is not contaminated by large numbers of protocluster galaxies. The 230\,arcmin$^{2}$ CF+ is therefore used as the control field in the rest of this work. 

A larger control field is required to determine the uncertainties due to the field-to-field variation of galaxy number counts. For this purpose we use the 0.8\,deg$^2$ UDS. Colour transformations are required to shift the observed UDS colours to HAWK-I colours since the UDS was observed with filters that have different passbands than HAWK-I filters. These transformations\footnote{\begin{eqnarray}
K_{HAWK-I}=K_{UDS}+0.09(H-K)_{UDS}-0.0635; \nonumber \\
(H-K)_{HAWK-I}=1.08(H-K)_{UDS}-0.138;\nonumber \\
(J-H)_{HAWK-I}=0.91(J-H)_{UDS}+0.0673. \nonumber
\end{eqnarray}
} were determined using stellar population models of high-redshift galaxies, and by matching the galaxy number counts and colour distributions of the \green\ and \black\ populations in the UDS to the CF+ (i.e., ensuring the UDS cumulative number counts in Fig.\,\ref{fig:comparison} match those of CF+). Therefore these transformations may not be suitable for sources with different colours, such as stars.

\subsection{Surface density of galaxies in the HzRG fields}
\label{sec:density}
\begin{figure}
 \begin{center}
 \includegraphics[width=0.95\columnwidth]{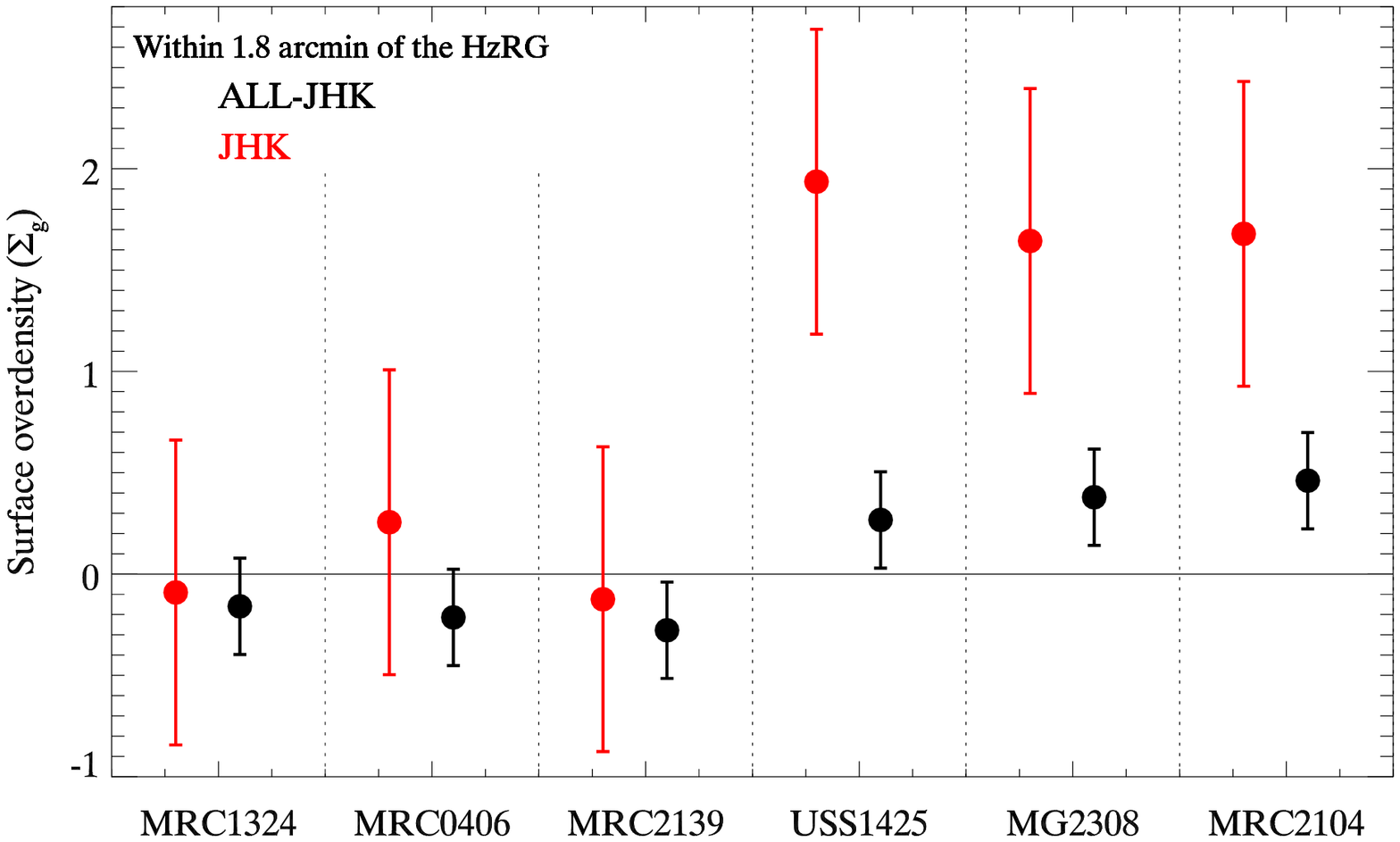}
  \includegraphics[width=0.95\columnwidth]{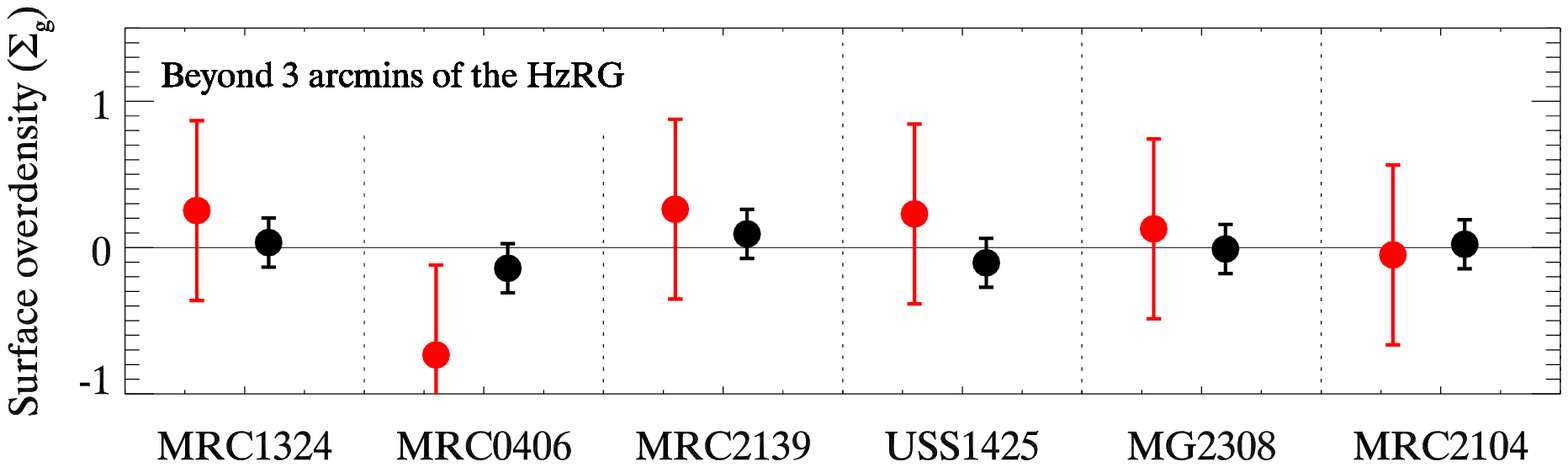}
\caption{ Surface overdensity within 1.5$\arcmin$ of the radio galaxies (top) and beyond 3$\arcmin$ (bottom). \green\ galaxies are plotted in green while \black\ galaxies are plotted in black.  The surface overdensity was measured using galaxies with  $K_{tot}<20.6$ and the HzRGs are not included. Three out of 6 fields contain significant overdensities in these two colour-selected populations.  None of the fields are overdense beyond 3$\arcmin$, which tells us that the overdensities are not due to zeropoint errors.  \label{fig:inner_outer}}
\end{center}
\end{figure}
\begin{table*}
  \centering
 \begin{tabular}{lcccccc}
  \hline 
   Field   & \MRCttf &\USS & \MRCofos &\MG&\MRCtoof&\MRCtotn \\
   \hline
  Type of galaxy         & && &&& \\
\black\ & $-0.2\pm0.2$ (65) &$0.3\pm0.2$ (91) &$-0.2\pm0.2$ (66) &$0.4\pm0.2$ (110) &$0.5\pm0.2$ (115) &$-0.3\pm0.2$ (58) \\
\green\ & $-0.1\pm0.7$ (2) & $1.9\pm0.8$ (6) & $0.3\pm0.7$ (3) & $1.6\pm0.7$ (6) & $1.7\pm0.7$ (6) & $-0.2\pm0.7$ (2)\\
\hline

Fraction of 1.8\arcmin radius UDS  & &&&&&\\
 cells as dense as HzRG field & 38\%&{\bf 0.3\%}&26\%&{\bf 0.4\%}&{\bf 0.3\%}& 43\% \\

\hline
\end{tabular}
   \caption{Galaxy overdensities of all colour-selected populations within 1.8$\arcmin$ of the 6 HzRGs. The number of galaxies in each sample are given in the parentheses. We do not include the HzRGs in these calculations. The probability of finding a 1.8$\arcmin$ radius cell in the UDS which is at least as dense in both colour-selected populations is given in the bottom row.  The overdensities around \USS, \MG\ and \MRCtoof\ deviate from the expected galaxy density by $\sim3\sigma$. \label{tab:overd}}
\end{table*}

The galaxy surface overdensity $\Sigma_g$ is the {\it excess} surface density of galaxies. It is calculated as  $\Sigma_g={(\Sigma_{obs}-\bar\Sigma)}/{\bar\Sigma}$, where  $\Sigma_{obs}$ is the observed surface density, and $\bar{\Sigma}$ is the expected surface density measured from the CF+ control field. 

The surface overdensity of the \green\ and \black\ galaxies was measured within 3\,co-moving Mpc (1.8$\arcmin$) of the HzRGs. This distance corresponds to approximately the virial radius of a massive local cluster. The fields were selected to include the HzRGs, so these galaxies were not included in either the \green\ or \black\ sample.   The results are shown in Fig.\,\ref{fig:inner_outer} and the number and overdensity of each population are listed in Table \ref{tab:overd}. The error bars represent 1$\sigma$ field-to-field variations in the surface density measured from 10,000 randomly positioned cells within the UDS.  The fields containing \USS, \MG\ and \MRCtoof\ are at least 1$\sigma$ overdense in both \green\ and \black\ galaxies. The other 3 fields show no significant overdensity in either population.

The surface overdensity of the \green\ and \black\ galaxies were also measured beyond 3$\arcmin$ of the HzRGs (see bottom panel of Fig.\,\ref{fig:inner_outer}). There are no significant overdensities in the outer regions of any field so the overdensities near the radio galaxies are not caused by zeropoint errors or inadequate subtraction of stars.  

$\MRCofos$ has a slight overdensity in \green\ galaxies within 1.5\arcmin\ and a 1$\sigma$ under-density in the outer regions.  Therefore the 1.8$\arcmin$ cell around the radio galaxy is significantly overdense in comparison to its local surroundings suggesting \MRCofos\ may have several nearby companions. 

\begin{figure*}
 \begin{center}
\includegraphics[width=0.66\columnwidth]{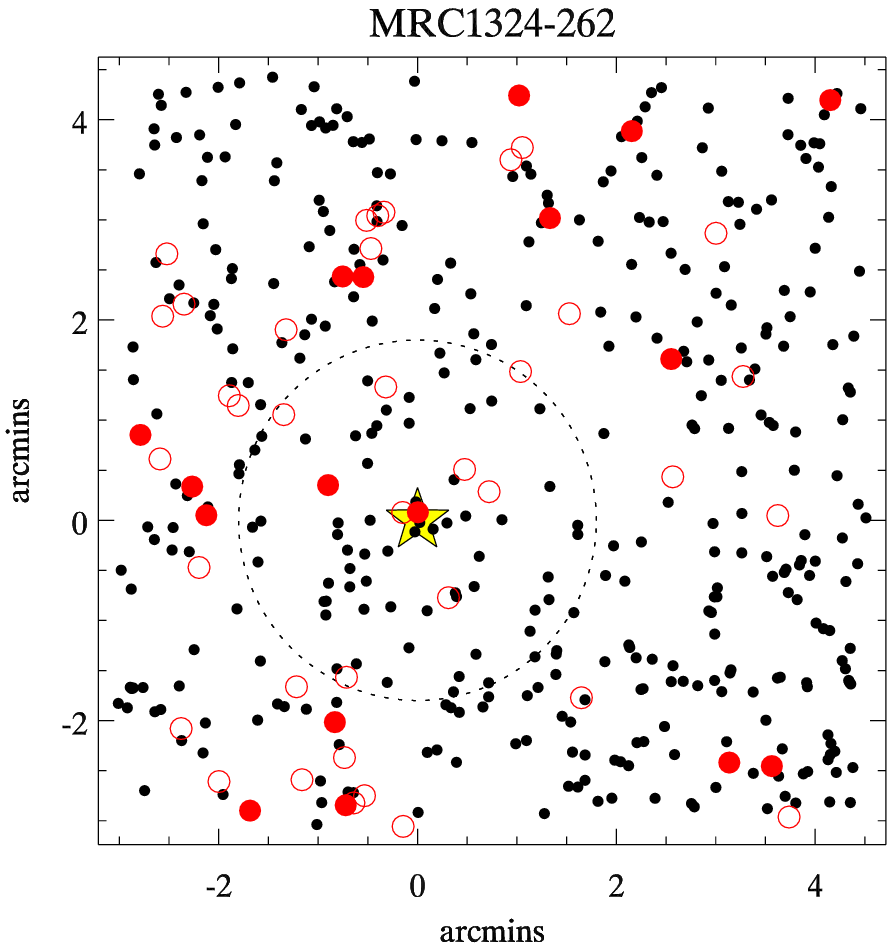}
  \includegraphics[width=0.66\columnwidth]{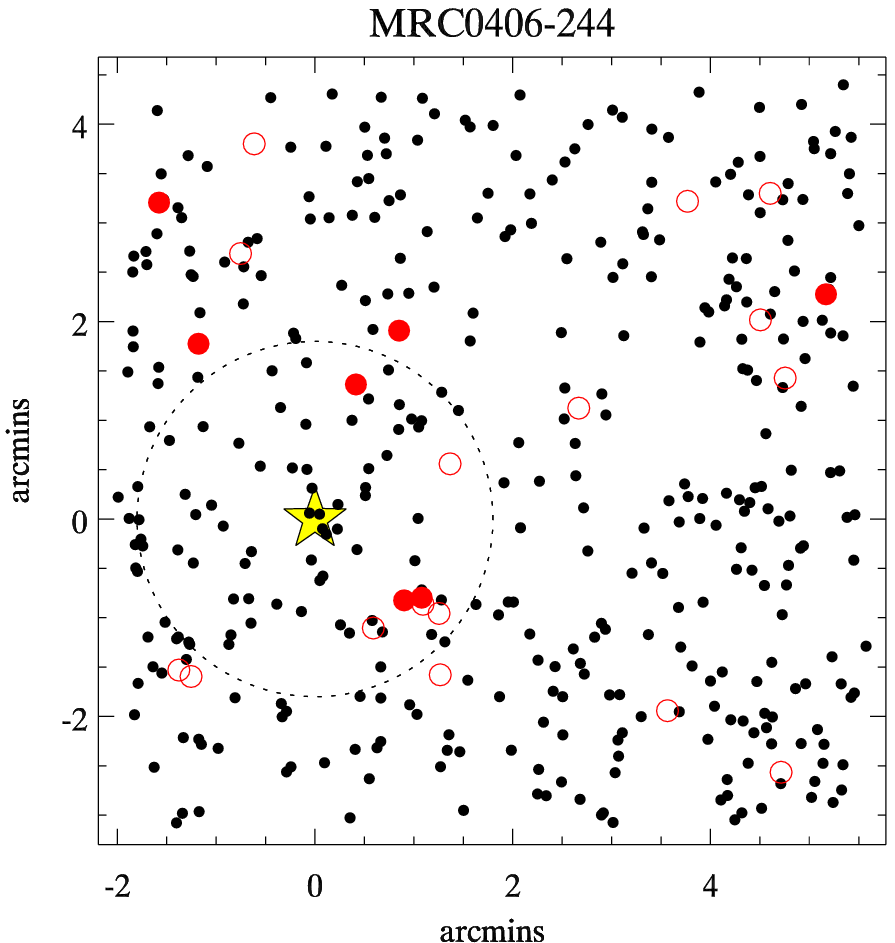}
   \includegraphics[width=0.66\columnwidth]{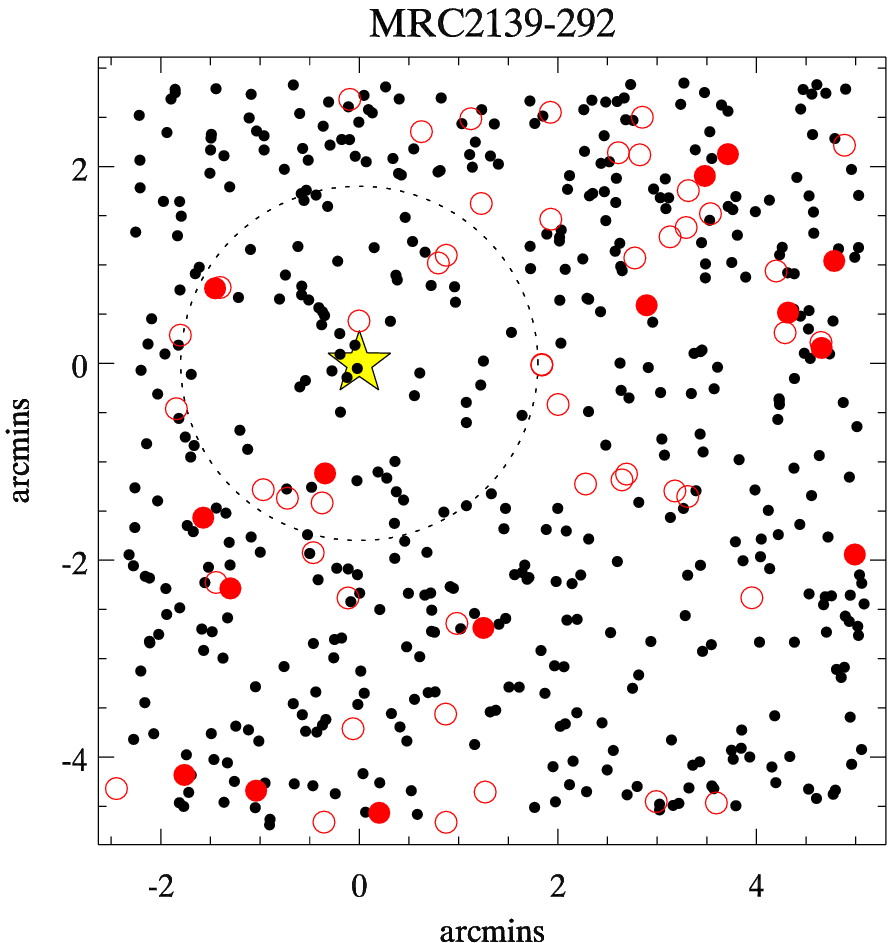}
    \includegraphics[width=0.66\columnwidth]{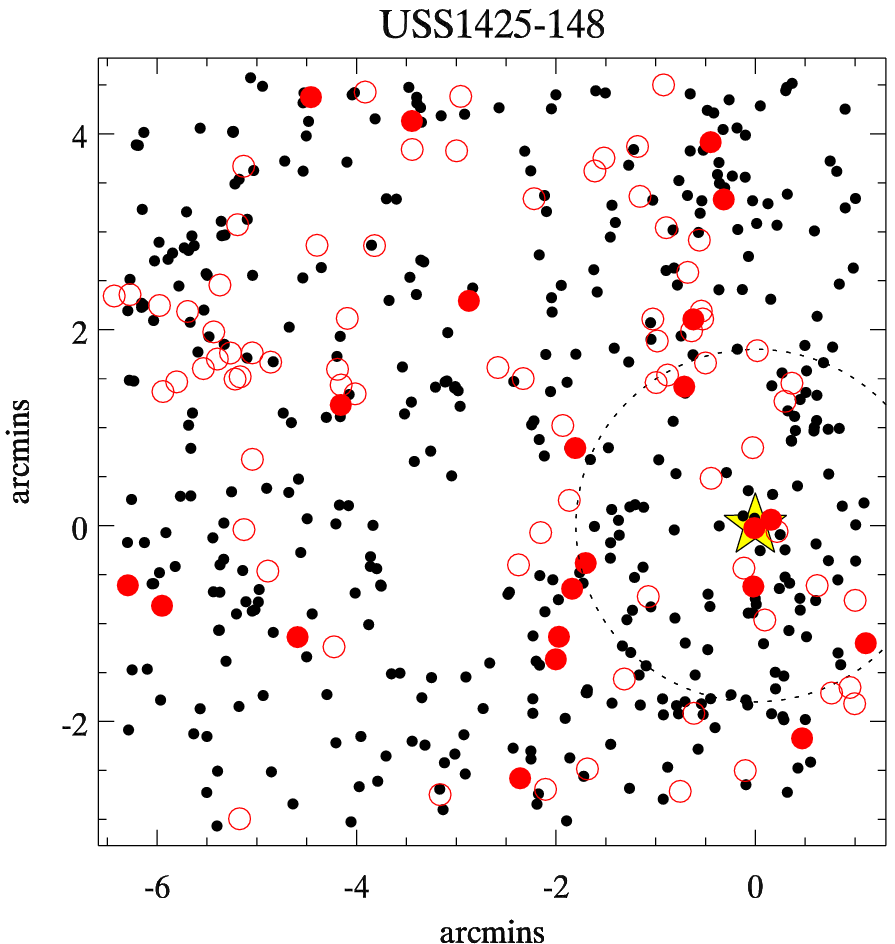}
     \includegraphics[width=0.66\columnwidth]{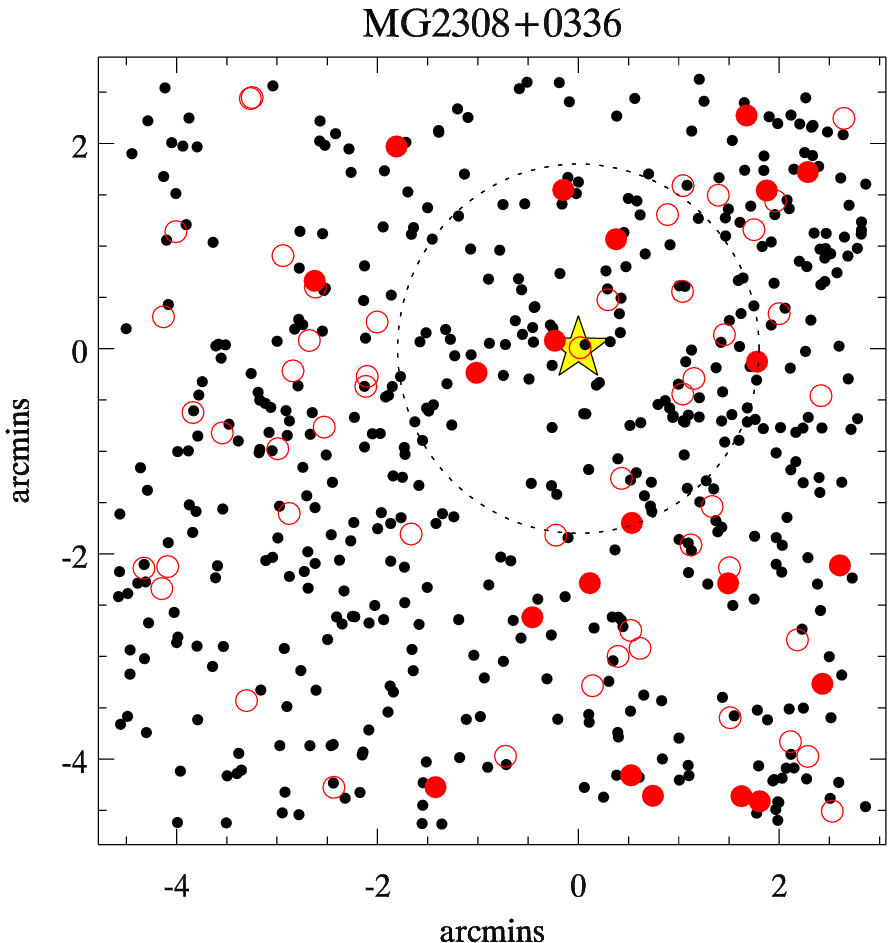}
      \includegraphics[width=0.66\columnwidth]{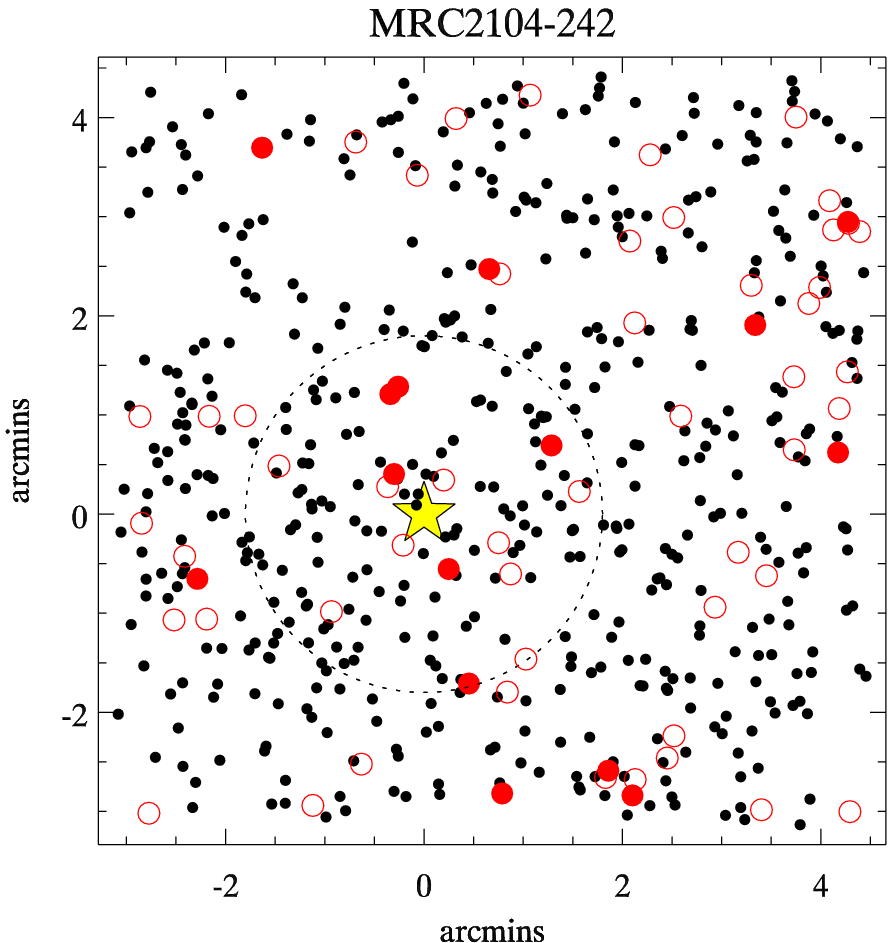}
\caption{Spatial distribution of the \green\ (red) and \black\ (black) galaxies in the 6 radio galaxy fields. Galaxies with $K_{tot}<20.6$ are marked as filled circles, while \green\ galaxies with  $20.6<K_{tot}<22.0$ are open circles. The yellow star marks the positions of the HzRGs and the dotted circles denotes 1.8$\arcmin$ from the HzRGs. The fields in the top row have no significant overdensity within 1.8$\arcmin$ of the radio galaxies, whilst the fields in the bottom row contain $>1\sigma$ overdensities of both \black\ and \green\ galaxies.  \label{fig:spatial_dist}}
\end{center}
\end{figure*}

The spatial distribution of \green\ and \black\ galaxies in the 6 radio galaxy fields are shown in Fig.\,\ref{fig:spatial_dist}. A visual inspection of these maps confirms that the fields containing \USS, \MG\ and \MRCtoof\ contain significantly more \black\ and \green\ galaxies than expected, whilst the 3 fields containing \MRCttf, \MRCofos\ and \MRCtotn\ do not.

\subsection{Significance of the surface overdensities}

The significance of the galaxy overdensity is the joint probability of finding an overdensity in both \green\ and \black\ populations. Since the populations are not mutually exclusive, the significance of each population cannot be simply combined. Instead 10,000 random 1.8$\arcmin$ radius cells within the 0.8\,deg$^{2}$ UDS were used to determine the probability of finding a cell which is as dense as the regions surrounding the HzRGs in both \black\ and  \green\ populations.

The probability of finding a 1.8$\arcmin$ radius cell within the UDS, with surface overdensities ($\Sigma_{obs}$) equal or greater than each of the radio galaxy fields is listed in the bottom row of Table \ref{tab:overd}. Between 26\% and 43\% of all UDS regions were as dense as the 1.8$\arcmin$ radius cells around \MRCttf, \MRCofos\ or \MRCtotn, so these fields do not contain more galaxies than expected. The probability of finding regions as dense as those surrounding \USS, \MG\ and \MRCtoof\  is $\sim0.3$\%, so the number of galaxies in these fields deviate from the expected galaxy density by $3\sigma$.

\section{Auto-correlation function analysis}
\label{clustering}
If the galaxies responsible for the overdensities in the 3 overdense HzRG fields are physically associated they should be strongly clustered. Whereas if the overdensity is caused by a chance line-of-sight alignment the clustering signal is not expected to be stronger than average. This clustering signal was measured using the auto-correlation function.

\subsection{Angular correlation function}
\begin{figure}
 \begin{center}
 \includegraphics[width=1\columnwidth]{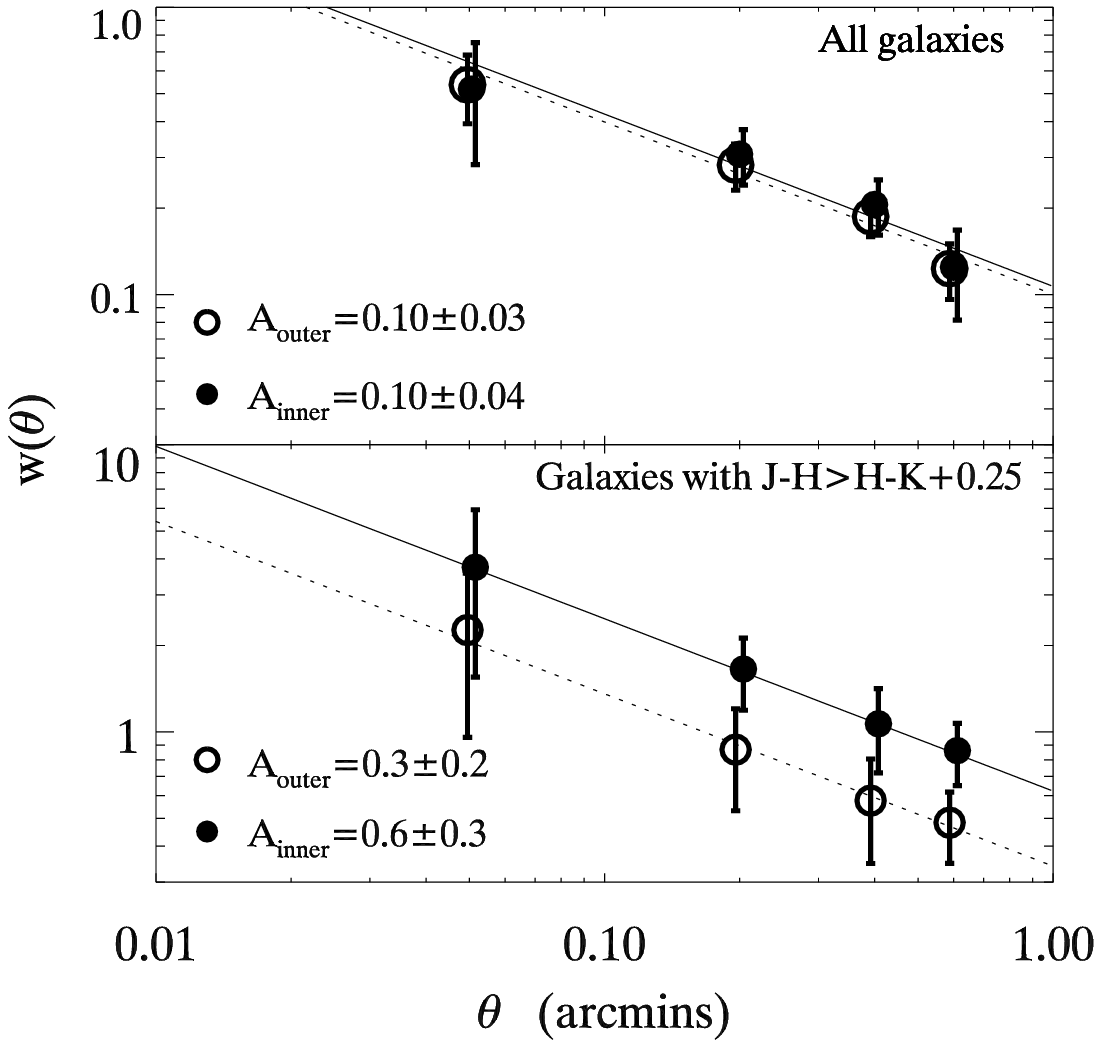}
\caption{The angular auto-correlation beyond 3$\arcmin$ (open circles and dotted line) and within 1.8$\arcmin$ (filled circles and solid line) of the 3 HzRGS in overdense fields. The angular correlation function of all galaxies near the HzRGs is consistent with further afield (top panel). The bottom panel shows the angular correlation function of a sample of galaxies that contains 1.5 times more galaxies near the HzRGs than in the outer regions. These galaxies are more strongly clustered close to the HzRGs, where there is a significant overdensity, than further afield, suggesting the galaxies responsible for the overdensity are physically associated.  \label{fig:wtheta}}
\end{center}
\end{figure}

The two-point angular correlation function, $w{\rm(}\theta{\rm)}$, is defined as the {\it excess} probability of finding two objects, in regions $\delta\Omega_1$ and $\delta\Omega_2$, separated by a distance $\theta$,
\begin{equation}
\delta P=\Sigma_{obs}^{2}[1+w(\theta)]\delta\Omega_1\delta\Omega_2, 
\end{equation}
where $\Sigma_{obs}$ is the surface density of galaxies in the field considered. Since $\Sigma_{obs}$ is taken into account, $w(\theta)$ will not be stronger in overdense regions just because there are more galaxies. We use the \citet{Landy1993} estimator for $w(\theta)$: 
\begin{equation}
w(\theta)=\frac{DD(\theta)-2DR(\theta)+RR(\theta)} {RR(\theta)}, 
\end{equation}
where $DD(\theta)$, $DR(\theta)$, and  $RR(\theta)$ are the number of data-data, data-random, and random-random pairs respectively, with angular separation between $\theta_i$ and $\theta_{i+1}$. 

To effectively sample the geometry of the detection area and take into account boundary effects, the random catalogues were constructed in the following way. The weight map image was divided into 4 regions of approximately equal effective exposure time.  The number of objects with $K_{tot}<20.6$\,mag were counted in each region, and the random catalogue constructed by placing 100 times as many objects at random locations in each of the 8 regions.  Terms involving random points were normalised such that $\Sigma_\theta DR(\theta)= \Sigma_\theta RR(\theta)= \Sigma_\theta DD(\theta)$. 

The angular correlation function was measured in 4 independent bins: 0.001 - 0.1\arcmin, 0.1-0.3\arcmin, 0.3-0.5\arcmin, and 0.5-0.7\arcmin. The uncertainty was estimated by bootstrap resampling \citep{Ling1986} and is quoted at the 1$\sigma$ level. To measure the clustering signal $w(\theta)$ was approximated by a power law, 
\begin{equation}
w(\theta)=A_{w}\theta^{-\beta}-{\rm IC}, 
\label{eqn:wtheta}
\end{equation}
where $A_w$ is the amplitude of the angular correlation function,  and $\beta$ is the slope.  We corrected for the integral constraint, which results from the finite field size,
using a constant, $IC$, determined from the random catalogues following the approach used by \citet{Infante1994} and \citet{Roche1999}:
\begin{equation}
IC\simeq \sigma_w^2=\frac{\Sigma_i A_w\theta_i^{-\beta}RR(\theta_i)}{\Sigma_i RR.(\theta_i)}.
\label{eqn:IC}
\end{equation} 
The slope was assumed to be $\beta=0.6$ \citep{Adelberger2005} as there is insufficient data to fit both the slope and the amplitude. The amplitude of the angular correlation function, $A_{w}$, was determined by fitting Eqn.\,\ref{eqn:wtheta}.

\subsection{Galaxy sample}
To test the angular correlation function analysis we measure $A_w$ for all galaxies with $17.5<K_{tot}<20.6$, both within 1.8\arcmin\ ($A_{\rm{inner}}$), and beyond 3\arcmin\ of \USS, \MRCtoof\ and \MG\ ($A_{\rm{outer}}$). Since the galaxies responsible for the overdensity make up a negligible fraction of this galaxy sample, any strong clustering signal of the excess galaxies will be masked, and we expect $A_{\rm{inner}}$ to be consistent with $A_{\rm{outer}}$.

To recover the clustering signal of the galaxies responsible for the overdensity we select a sample of galaxies which contains a large excess. All 3 overdense fields contain $\sim3$ times the expected density of \green\ galaxies, but there are too few galaxies in this sample to obtain a measure of $w(\theta)$ in 4 independent $\theta$ bins.  We therefore select galaxies that meet a similar, but slightly more relaxed colour criterion ($J-H>H-K+0.25$).  58 galaxies with such colours lie within 1.8\arcmin\ of the 3 overdense HzRGs (a number density of 2\,arcmin$^{-2}$), compared to 279 in the CF+ (1.2\,arcmin$^{-2}$), indicating an overdensity of 0.7. Therefore the excess galaxies make up 40\% of this sample. 

We measured the amplitude of the angular correlation function of galaxies with $J-H>H-K+0.25$ colours within 1.8\arcmin\ of \USS, \MRCtoof\ and \MG\  ($A_{inner}$). This was compared to the $A_w$ of galaxies with the same colours and magnitudes that lie beyond 3\arcmin\ of the same HzRGs ($A_{\rm{outer}}$). In this sample the clustering signal of the excess galaxies should not be strongly masked. If the excess galaxies are physically associated, we expect $A_{\rm{inner}}$ to be greater than $A_{\rm{outer}}$.

\subsection{Results}

The auto-correlation functions are plotted in Fig.\,\ref{fig:wtheta} for all galaxies (top panel) and galaxies with $J-H>H-K+0.25$ (bottom panel). The auto-correlation function of the galaxies beyond  3$\arcmin$ of the HzRGs and within 1.8$\arcmin$  are plotted as open and solid circles respectively, and the amplitudes $A_{\rm outer}$ and $A_{\rm inner}$ are given in the legend. 

As expected, $A_w$ of all galaxies near the 3 overdense HzRGs is consistent with that measured in the outer regions, $A_{\rm outer}=A_{\rm inner}=0.1$. The angular correlation function of both regions agree at all scales, suggesting there is no bias due to the different regions, and that $w(\theta)$ and $A_{w}$ for the overdense sample may be robustly compared without fear of systematic bias. 

Galaxies with $J - H > H - K + 0.25$ colours that lie within 1.8\arcmin\ of the HzRGs have larger $w(\theta)$ at all scales compared to the outer region of the fields.  $A_{\rm inner}$ is 2 times greater than $A_{\rm outer}$, however the uncertainties are large due to the small number of galaxies in this galaxy sample, and this result is significant to only 1$\sigma$. Nonetheless the galaxies responsible for the overdensity are more strongly clustered near the HzRGs than similarly selected galaxies at larger radii. This suggests the galaxies responsible for the overdensity are physically associated with one another.  

We have checked our results assuming $\beta=0.8$, which is also commonly used in the literature. We find that $A_{w}$ is reduced by a factor of approximately 2, but the disparity between $A_{\rm outer}$ and $A_{\rm inner}$ is still present for galaxies with $J - H > H - K + 0.25$ colours.

\section{Properties of the protocluster galaxy candidates}
\label{stat}
In this section we aim to measure the properties of {\it all} protocluster galaxy candidates in the 3 overdense HzRG fields. 

Only a small fraction of protocluster galaxy candidates will be in the \green\ sample, as this criterion selects less than 15\% of galaxies at $2.2<z_{phot}<2.7$. Thus the \green\ galaxies are not representative of the entire galaxy population surrounding the HzRGs. 

Most of the $z\sim2.4$ protocluster galaxy candidates are expected to be \black\ galaxies, as this sample encompasses 98\% of galaxies with $2.2<z_{phot}<2.7$. However the overdensity of \black\ galaxies in the 3 overdense HzRG fields is $\sim0.4$, so more than 70\% of \black\ galaxies lie in the foreground or background. 
This interloper population must be removed from the \black\ sample to reveal the properties of the protocluster galaxy candidates galaxies.

\subsection{Deriving a flux-limited sample of protocluster galaxies through statistical subtraction}
We do not know which particular \black\ galaxy is an interloper, but we can statistically subtract the interloping galaxies from the \black\ sample. 

To derived the statistical galaxy excess, the number density, magnitudes and colours of the interlopers were measured from the control field (CF+),  and subtracted from the galaxy counts of the 3 overdense HzRG fields.  If the galaxy overdensities are due to protoclusters, this method should ideally remove all foreground and background galaxies, to leave only the protocluster galaxies. 

A major advantage of this method is that, by construction, a flux-limited sample of candidate protocluster galaxies is selected, irrespective of colour or star formation history. This allows us to determine the colours and magnitudes of {\it all} protocluster members.

Accurate statistical subtraction requires both the control field (CF+)  and the overdense fields be complete to the same magnitude. Thus the galaxy catalogues are limited to the 90\% completeness limit of the lowest-exposure time region in the shallowest field ($K_{tot}<20.6$\,mag), and the derived sample of excess galaxies are complete to this flux limit.

\subsubsection{Uncertainties and biases}

The greatest uncertainty in the statistical subtraction method comes from cosmic variance of the interloping galaxy population. To quantify the uncertainty, we ran 1000 Monte Carlo simulations using the UDS, by selecting and combining 3 random 2.1\arcmin\ radius regions and subtracting the interloping galaxies as measured from a randomly positioned 230\,arcmin$^2$ region. The standard deviation of the resulting statistical excess (or deficit) gives the 1$\sigma$ uncertainty in the removal of the interloping population. When deriving the uncertainty on a galaxy property, such as colour, galaxies were divided in colour bins and the uncertainty is the standard deviation within each colour bin. 

To check that selecting \black\ galaxies does not bias the results, we tested the statistical subtraction method using all galaxies with $K_{tot}<20.6$. The derived properties of the excess galaxies are fully consistent with the following presented results, which suggests that few, if any, protocluster galaxy candidates are removed by making the \black\ selection.

\label{sec:uncert}

\subsection{Radial extent and volume of the overdensities}
\label{radius}
\begin{figure}
 \begin{center}
 \includegraphics[width=1\columnwidth]{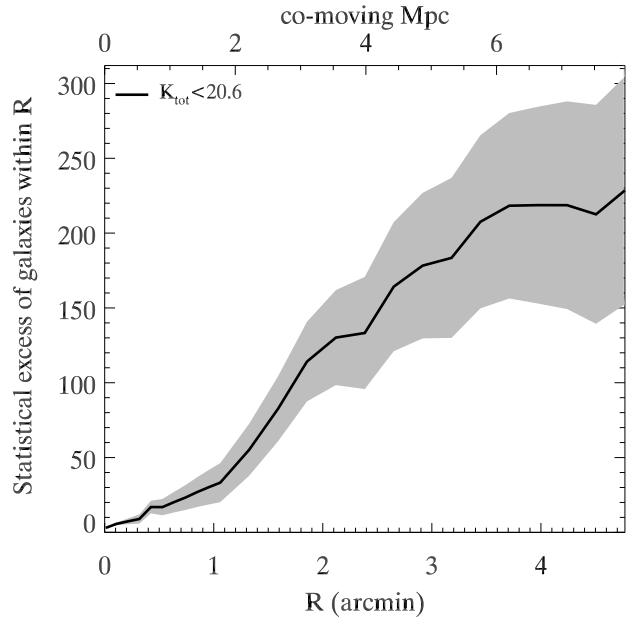}
 \caption{The cumulative number of excess galaxies with $K_{tot}<20.6$ within a projected distance R of the 3 HzRGs in overdense fields (\USS, \MRCtoof\ and \MG).  The radial extent of the average overdensity is located where the cumulative distribution stops increasing. Approximately 220 excess galaxies are situated within 3.5$\arcmin$ (6 co-moving Mpc) of the 3 radio galaxies. The {\it average} protocluster contains 1/3 of the shown numbers. The grey shaded area shows 1$\sigma$ field-to-field variations. \label{fig:number_of_excess_gal_withinR}}
\end{center}
\end{figure}

To determine the radial extent of the overdensity, we measure the statistical excess of \black\ galaxies within a projected radius, R,  of the HzRGs.  This was achieved by counting \black\ galaxies within a radius R of \USS, \MRCtoof\ and \MG, and subtracting the expected number of \black\ galaxies within this area (determined from CF+). The radial extent of the overdensity is marked where the cumulative distribution stops increasing. 

By inspection of the cumulative distribution in Fig.\,\ref{fig:number_of_excess_gal_withinR}, we find that the {\it average} overdensity extends to a projected radius of $\sim$3.5$\arcmin$ (6~co-moving~Mpc at $z=2.4$) and contains approximately 75$\pm$20 galaxies with $K_{tot}<20.6$.  Half of the statistical excess lies within 1.8\arcmin\ (3 co-moving Mpc) of the HzRGs, so this central region is twice as dense as the entire overdensity.

Assuming the overdensity lies in a cube of side 12\,Mpc, the co-moving of the overdensity is $\sim1700$\,Mpc$^3$. Dark matter simulations of the early Universe show that matter collapses in filaments, and our selection method preferentially selects filaments that are close to being perpendicular to the plane of the sky. Therefore this estimate of the volume is very uncertain.

\begin{figure}
 \begin{center}
 \includegraphics[width=1\columnwidth]{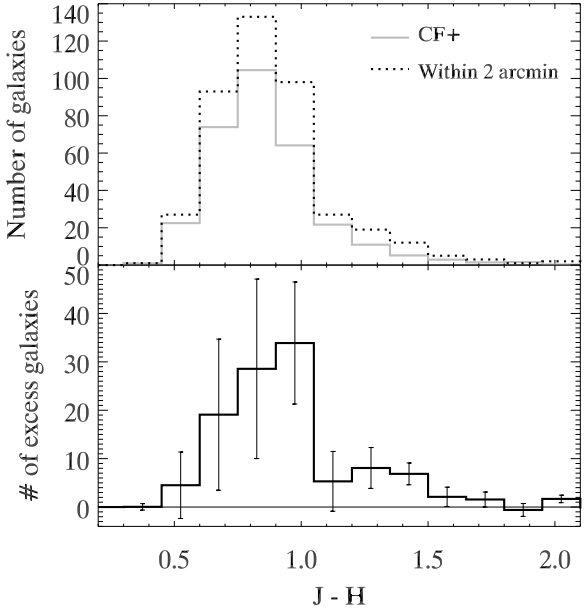}
 \caption{The top panel shows the $J-H$ colour distribution of galaxies in the 3 overdense fields (dashed histogram) and CF+ galaxies, normalised to the same area (grey histogram). The difference between these two distributions reveal the $J-H$ colours of the excess galaxies, which is shown in the bottom panel in the solid black histogram.  A two-sided Kolmogorov-Smirnov test comparing the colours of the excess galaxies to the CF+ galaxies results in prob$=0.001$, implying the colour distribution of the two groups of galaxies differ significantly. The excess galaxies are therefore not randomly drawn from all redshifts, but probably comprise a group of galaxies at a particular redshift. \label{fig:JH_colour}}
\end{center}
\end{figure}
\subsection{Colours of the excess galaxies}
\label{properties}

The colour distribution of the statistical excess was derived by subtracting the colour distribution of the interloping \black\ galaxies (measured from the CF+ and normalised by area) from the colour distribution of \black\ galaxies within  the 3 HzRGs in overdense fields. 

Ideally the colours of all excess galaxies within the HzRG fields should be derived. However, the variance due to the interloper population is approximately proportional to the field area. Therefore the colours of the statistical excess was only measured within 3.5 co-moving Mpc (2.1\arcmin) of \USS, \MG\ and \MRCtoof, where the number of excess galaxies is more than 4 times the uncertainty (see Fig.\,\ref{fig:number_of_excess_gal_withinR}).  

 \begin{figure*}
 \begin{center}
  \includegraphics[width=2\columnwidth]{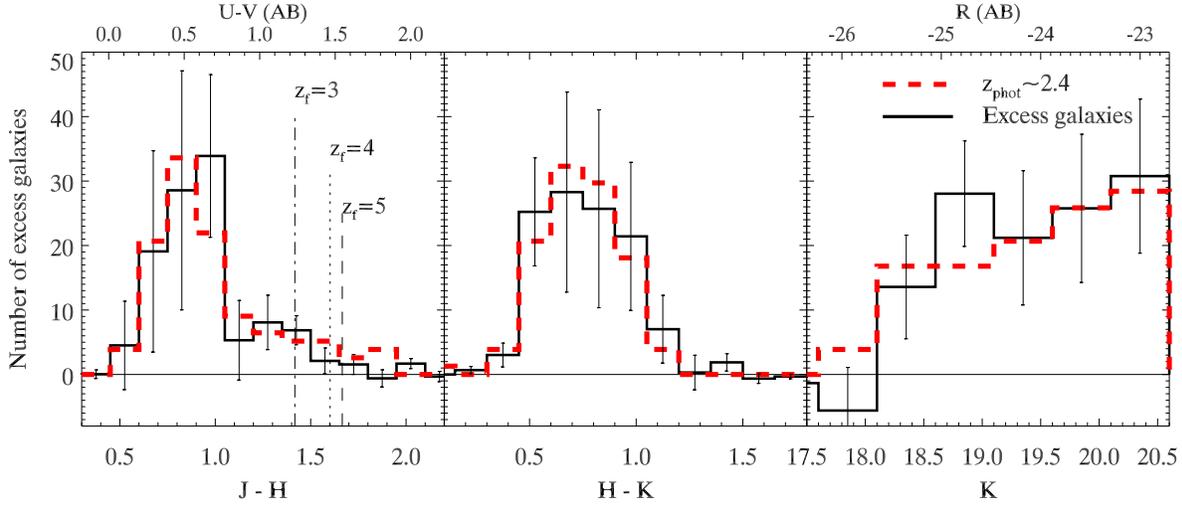}
 \caption{The observed near-infrared colours and \ktot\ fluxes of the statistical galaxy excess (solid black histograms) compared to those of the $z_{\rm phot}$-UDS sample (dashed red histogram). $z_{\rm phot}$-UDS galaxies are selected from the UDS to have (i) $2.2<z_{phot}<2.7$, (ii) $\log\,\chi^2<2.9$, and (iii) the 3$\sigma$ confidence intervals of the redshift estimate lie within the redshift range $2.2<z<2.7$. There are 87 galaxies in the $z_{\rm phot}$-UDS sample, and the distributions have been normalised to the number of excess galaxies.The colours and masses of $z_{\rm phot}$-UDS and excess galaxies are similar with KS probability of p=0.82. The red tail seen in both $J-H$ colour distributions is unique to galaxies at $2.2<z<2.7$, whose redshifted Balmer break shifts between the $J$ and $H$ passbands. The $J-H$  colour distribution is bimodal, with a strong peak of blue galaxies and an extended tail of red galaxies. There is no prominent peak of red galaxies. The expected positions of the $z=2.4$ red sequence are marked as dot-dashed, dotted, and dashed lines assuming the galaxies formed at $z_{f}=3, 4, 5$ respectively.  Most protocluster galaxies are blue so they are still forming stars at $z\sim2.4$, and the red galaxies are consistent with a formation redshift of $z_f\sim3$\label{fig:compare_UDS}}
\end{center}
\end{figure*}

\subsubsection{Comparing the colours of the excess galaxies to control field galaxies} 

 Fig.\,\ref{fig:JH_colour} displays the colour distribution of \black\ galaxies in the CF+, galaxies within 2.1\arcmin\ of the 3 overdense HzRGs, and the derived statistical galaxy excess. The colour distribution of the statistical excess differs from the control field, which is quantified using a two-sided Kolmogorov-Smirnov (KS) test. 

The KS test requires continuous unbinned data, so a mock catalogue of galaxy colours was created by sampling random values whilst constraining the catalogue to match the observed colour distribution of the statistical excess. The KS test was performed 1000 times and in each test the colours of the excess galaxies were drawn from a different set of random values. The probability, p,  gives the significance level of the test (ranging between 0 and 1) and is the median of all derived p. 

The KS test rejects the null hypothesis with p=0.001, so the colour distribution of the excess galaxies differs significantly from the colour distribution of galaxies in the CF+. The excess galaxies in the 3 overdense fields are not drawn randomly from the control field population and are therefore unlikely to be caused by a chance line-of-sight alignment. 

  \subsubsection{Comparing the colours of the excess galaxies to $z_{\rm phot}\sim2.4$ galaxies} 
 \label{UDS_z_gals}

The colour distribution of UDS galaxies with photometric redshifts in the range $2.2<z_{phot}<2.7$ ($z_{\rm phot}-$UDS sample; see Section \ref{UDS_phot_section}) is shown in Fig.\,\ref{fig:compare_UDS}. The distribution is similar to that of $2<z_{\rm phot}<3.5$ galaxies in \citet{Wuyts2007} which span similar fluxes\footnote{ ($U-V$)$_{\rm ~rest}{\rm(AB)}=1.26\times{\rm (}J-H{\rm )}-0.558$ for galaxies at $z\sim2.4$}. Both distributions peak at $U-V\sim0.5$, and there are relatively few galaxies redder than $U-V=1.0$, or bluer than $U-V=0.0$ (see Fig.\,8 in \citealt{Wuyts2007}). The $z_{\rm phot}$-UDS sample is bluer than the $2<z<3$ galaxies of \citet{vanDokkum2006}, but their study is limited to galaxies with stellar masses greater than $10^{11}$\Msun\ which are generally brighter than our flux-limited sample. We conclude that the colour distribution of the $z_{\rm phot}$-UDS sample is a good representation of $z\sim2.4$ galaxy colours.

The $J-H$ and $H-K$ colours, and fluxes of galaxies comprising the statistical excess are compared to the UDS galaxies with $2.2<z_{\rm phot}<2.7$ in Fig.\,\ref{fig:compare_UDS}. The flux distributions of both samples are consistent, so a robust comparison of the two samples can be made. 

The colour distributions of the excess galaxies are very similar to the UDS $z_{\rm phot}\sim2.4$ galaxies.  The red tail in the $J-H$ colour distribution is unique to galaxies at $2.2<z<2.7$. It is caused by Balmer break of galaxies at this redshift lying between the $J$ and $H$ passbands. Both the excess galaxies and the $z_{\rm phot}\sim2.4$ galaxies have this red tail in the $J-H$ colour distribution. A KS test does not reject the null hypothesis that the excess galaxies and the UDS $z_{\rm phot}\sim2.4$ galaxies are drawn from the same underlying distribution (p=0.82). Since the excess galaxies have similar $J-H$ and $H-K$ colours, and magnitudes to $z\sim2.4$ galaxies, it is likely that they are situated at the same redshift as the HzRGs.

\section{Discussion}
\label{discussion}
The 3 HzRGs \USS, \MG\ and \MRCtoof, are surrounded by galaxy surface overdensities that deviate from the expected density by $3\sigma$. The excess galaxies responsible for the overdensities are strongly clustered, suggesting they are physically associated. One of the strongest pieces of evidence that the candidate protocluster galaxies are associated with the radio galaxies is the similarity between the colours of the candidate protocluster galaxies and $z_{phot}\sim2.4$ galaxies in the $K-$selected Ultra-Deep Survey (see Fig.\,\ref{fig:compare_UDS}). It would be a strange coincidence that the colour and luminosity distributions match so well if both these samples did not consist of $z\sim2.4$ galaxies, especially considering they were selected by different methods. In this section we examine the properties of these overdensities.

\subsection{Evolution of the overdensities around $z\sim2.4$ HzRGs}

\label{mass_size}

\subsubsection{Average mass enclosed in the overdensities}
Assuming all the mass within the volume collapses into a single virialized structure, the mass of the collapsed structure is estimated by  
\begin{equation}
{\rm M}=\bar\rho {\rm V}(1+\delta_{m})
\label{eqn:mass}
\end{equation}
\citep{Steidel1998} where V is the volume of the overdensity, $\bar\rho$ is the average density of the Universe ($1.9\times10^{10}$\Msun Mpc$^{-3}$), and $\delta_{m}$ is the mass overdensity, related to the galaxy overdensity through the bias parameter $\delta_m=\delta_g / b$. 

To determine the overdensity of galaxies, we compare the number density of galaxies in the overdensities to the $z=2.5$ field. \citet{Wuyts2009} measure the density of $z\sim2.4$ galaxies with M$>4\times10^{10}$\Msun\ to be $3.3\times10^{-4}$Mpc$^{-3}$.  
Stellar masses of the excess galaxies were estimated from \ktot\ using the same \citet{BC03} stellar synthesis models as \citet{Wuyts2009}. Six stellar synthesis models (plotted in Fig.\,\ref{fig:JH_HK_track}) were redshifted to $z=2.35$, 2.46 and 2.49 (i.e., the redshifts of the radio galaxies). Moderate amounts of dust extinction (up to $A_V$=1\,mag) were applied to the models, and the models were convolved with the HAWK-I $J,\,H$ and $Ks$ filter bandwidths to obtain the following relation:
$\log~ M_*=[K_{tot}-43.32\,(\pm0.15)-2.79\,(\pm0.16)\times(J-H)]/-2.5$.
The uncertainties are derived from the 3$\sigma$  scatter about the linear relation.

Some galaxies responsible for the overdensity have $J-H\sim1.5$ so the flux-limited sample is only complete to a mass-limit of $\sim6\times10^{10}$\Msun. Therefore the number density of M$>4\times10^{10}$\Msun\ galaxies in the HzRG fields, and the mass estimates of the overdensities, are lower limits.

The number density of excess galaxies with M$>4\times10^{10}$\Msun\ within 6\,Mpc of the 3 overdense HzRGs is $24\pm12 / {\rm V}$\,Mpc$^{-3}$.  The uncertainties quoted include uncertainties in the mass estimation and the statistical subtraction of the interlopers. Assuming $V\sim1700$\,Mpc$^3$, the galaxy overdensity is $\delta_g\sim42\pm21$. The bias parameter for galaxies with a number density of $3.3\times10^{-4}$Mpc$^{-3}$ at $z\sim2.4$ is $\sim3$ \citep{Quadri2007}, thus the mass overdensity is $\delta_m=14\pm7$.

The mass contained within the average radio galaxy overdensity is $\sim2-7\times10^{14}$\Msun\ which is comparable to the mass of present-day clusters of galaxies, and is consistent with the mass estimates of other HzRG-selected protoclusters \citep{Venemans2007}.

The volume of the overdensity is very uncertain (see Section \ref{radius}), but this does not affect the estimation of mass enclosed within the overdensity, as the volume in Equation\,\ref{eqn:mass} cancels with the volume used to calculate $\delta_m$. 

\subsubsection{Collapse time}
In a matter dominated Universe, the time for a uniform sphere to separate from the Hubble flow and collapse is $\sim(1/\,{6\pi \rm G}\,\rho)^{1/2}$ \citep{Peacock1999}. The collapse time for these cluster-sized overdensities is $\sim6$\,Gyr. This  estimate is very uncertain as the volume of the overdensity is poorly constrained, but this short collapse time implies these overdensities may virialize by the present day. 
\\

In conclusion, the 3 HzRG fields with significant projected surface overdensities fulfil the main physical requirements for evolving into a galaxy cluster or group by the present day.

\subsection{Rest-frame $U-V$ colours of galaxies in $z>2$ protoclusters}

The rest-frame $U-V$ colours of the flux-limited sample of protocluster candidates are derived through ($U-V$)$_{\rm ~rest}{\rm(AB)}=1.26\times{\rm (}J-H_{\rm observed}{\rm )}-0.558$ and shown in Fig.\,\ref{fig:compare_UDS}. The colour distribution is bimodal, consisting of a dominant blue sequence, and a poorly populated red tail.  $77\pm10$\% of the protocluster galaxies are blue with $(U-V)_{\rm AB}<0.85$. However, this is an upper limit as blue galaxies are detected to lower masses than red galaxies in the flux-limited sample. 

The blue protocluster candidates have a typical colour of $U-V_{\rm AB}\sim0.6$\,mag, which is bluer than local spiral galaxies, and typical of a present-day irregular galaxy. Such blue colours suggest that these galaxies are forming stars.

The colours of the galaxies in the red tail of the distribution are consistent with that of a cluster red sequence which formed the majority of its stars at $z_f\sim3$, a mere $0.5$\,Gyr before the epoch of observation. However, the red sequence is poorly populated compared to those observed in clusters at $z\lesssim1.6$ \citep[e.g.,][]{Blakeslee2003,Papovich2010,Tanaka2010}. The errors on the observed colours and the slightly different redshifts of the HzRGs may increase the scatter of the galaxies on the red sequence.

A colour bimodality, and prominent red sequence has been found up to $z\sim2.5$ using large area photometric surveys \citep[e.g.,][]{Williams2009a,Brammer2009}. These surveys probe much larger volumes than the protoclusters, and detect greater numbers of massive galaxies that populate the red sequence. The lack of a strong red sequence in the 3 protocluster candidates may be due to the low number of massive galaxies within each protocluster.

No significant difference is observed between the colours of the protocluster galaxies and field galaxies at $z_{phot}\sim2.4$. This suggests that global environmental effects have yet to become important in clusters by $z\sim2.4$. However, the samples of protocluster and field galaxies are selected by different methods, so we refrain from making detailed comparisons between these two populations. 

\citet{Steidel2005} found a significant fraction of the galaxies in a $z=2.3$ protocluster were detected in the rest-frame UV. Hence these galaxies were still forming stars and were relatively blue at this redshift, which is consistent with our findings. Through detailed SED modelling, the ages and stellar masses of the $z=2.3$ protocluster galaxies were found to be a factor of 2 greater than field galaxies. While our data does not allow us to perform a similar analysis, future follow-up observations may reveal how wide-spread such evolutionary trends are among samples of protoclusters.

Our results show that the majority of galaxies in the $z>2$ protocluster candidates have colours typical of star-forming galaxies. The colour, low scatter and non-evolving slope of the tight red sequence in lower-redshift clusters imply the stars that constitute cluster ellipticals formed contemporaneously at $z>2$ \citep{Ellis1997,Gladders1998,Mei2009}. The scatter and colour of the red sequence in a $z=1.62$ cluster indicates this epoch of intense stellar-buildup occurred at $z=2.4\pm0.15$  \citep{Papovich2010}, whilst the evolution in the mass-to-light ratio of elliptical galaxies up to $z=1.3$ imply a formation redshift of $2-2.5$ \citep{vanDokkum2007}. The  $z>2$ protocluster candidates may have been caught in this starburst phase, during which they form the bulk of the stellar population that will later reside in their elliptical galaxies.

\section{Conclusions}
\label{conclusions}

We have conducted a near-infrared survey of the environment of 6 high redshift radio galaxies (HzRG) at $2.2<z<2.7$. We use colour-cuts to identify galaxies in this redshift interval, and find excess numbers of these galaxies around 3 HzRGs:  \USS, \MRCtoof\ and \MG.

Within 1.8\arcmin\ (3 co-moving Mpc) of these 3 HzRGs approximately $3$ times more colour-selected galaxies are observed than expected. The surface density of galaxies with colours of $2.2<z<2.6$ galaxies deviate by $3\sigma$ from the field density. The clustering of the excess galaxies responsible for the overdensity was measured through the angular auto-correlation function. The excess galaxies are more strongly clustered than similarly-selected galaxies in the control field, suggesting they are physically associated.

We derive a flux-limited sample of protocluster galaxy candidates by statistically subtracting the fore- and background sources from the 3 overdense fields. The observed colour distribution of the statistical galaxy excess is consistent with that of galaxies with photometric redshifts in the interval $2.2<z<2.7$, suggesting that the excess galaxies are located in a narrow redshift range near the HzRGs, and are therefore likely to be protocluster galaxies associated with the HzRGs. However, spectra of the protocluster galaxy candidates are required to confirm whether the overdensities are coherent structures.

The large-scale structure around the HzRGs extends approximately 12\,Mpc (co-moving), and contains $75\pm20$ galaxies with $K_{tot}<20.6$\,mag. The average mass contained within these structures exceeds $10^{14}$\Msun, and these dense regions may collapse and virialize by $z\sim0$. Thus these objects may evolve into groups or clusters of galaxies by the present day. 

We find the rest-frame $U-V$ colour distribution of the flux-limited sample of protocluster galaxies is bimodal, consisting of a dominant blue sequence and a poorly populated red sequence. 77$\pm10$\% of protocluster galaxies have blue colours, typical of star-forming galaxies. This implies that the bulk of stars that constitute cluster ellipticals have not  formed by $z\sim2.4$, or are too blue to lie on the red sequence. The $U-V$ colours of the protocluster member candidates are consistent with a formation redshift of $z\lesssim3$.

\section{Acknowledgments}
We would like to thank Rik Williams and Ryan Quadri for their UDS catalogues, and Masa Tanaka for useful discussions. We thank the referee, whose suggestions greatly improved the paper. 

NAH is supported by a Postdoctoral Fellowship from the Science and Technology Facilities Council (ST/G004994/1), and an Anne McLaren Fellowship from the University of Nottingham. JK thanks the DFG for support via German-Israeli Project Cooperation grant STE1869/1-1.GE625/15-1.

This research has been based on observations made with the VLT at ESO, Paranal, Chile, programs 081.A-0110(A), 0.82.A-0738(C), 0.83.A-0231(A).
\bibliographystyle{mn2e}\bibliography{cluster_progenitors,mn-jour}

\begin{thebibliography}{}

\bibitem[\protect\citeauthoryear{{Adelberger}, {Steidel}, {Pettini}, {Shapley},
  {Reddy} \& {Erb}}{{Adelberger} et~al.}{2005}]{Adelberger2005}
{Adelberger} K.~L.,  {Steidel} C.~C.,  {Pettini} M.,  {Shapley} A.~E.,  {Reddy}
  N.~A.,    {Erb} D.~K.,  2005, \apj, 619, 697

\bibitem[\protect\citeauthoryear{{Bertin} \& {Arnouts}}{{Bertin} \&
  {Arnouts}}{1996}]{Bertin1996}
{Bertin} E.,  {Arnouts} S.,  1996, \aaps, 117, 393

\bibitem[\protect\citeauthoryear{{Best}, {Lehnert}, {Miley} \&
  {R{\"o}ttgering}}{{Best} et~al.}{2003}]{Best2003}
{Best} P.~N.,  {Lehnert} M.~D.,  {Miley} G.~K.,    {R{\"o}ttgering} H.~J.~A.,
  2003, \mnras, 343, 1

\bibitem[\protect\citeauthoryear{{Blakeslee}, {Anderson}, {Meurer},
  {Ben{\'{\i}}tez} \& {Magee}}{{Blakeslee} et~al.}{2003}]{Blakeslee2003}
{Blakeslee} J.~P.,  {Anderson} K.~R.,  {Meurer} G.~R.,  {Ben{\'{\i}}tez} N.,
  {Magee} D.,  2003, in {Payne} H.~E.,  {Jedrzejewski} R.~I.,   {Hook} R.~N.,
  eds, Astronomical Data Analysis Software and Systems XII Vol.~295 of
  Astronomical Society of the Pacific Conference Series, {An Automatic Image
  Reduction Pipeline for the Advanced Camera for Surveys}.
pp 257--+

\bibitem[\protect\citeauthoryear{{Bower}, {Lucey} \& {Ellis}}{{Bower}
  et~al.}{1992}]{Bower1992}
{Bower} R.~G.,  {Lucey} J.~R.,    {Ellis} R.~S.,  1992, \mnras, 254, 589

\bibitem[\protect\citeauthoryear{{Brammer}, {Whitaker}, {van Dokkum},
  {Marchesini}, {Labb{\'e}}, {Franx}, {Kriek}, {Quadri}, {Illingworth}, {Lee},
  {Muzzin} \& {Rudnick}}{{Brammer} et~al.}{2009}]{Brammer2009}
{Brammer} G.~B.,  {Whitaker} K.~E.,  {van Dokkum} P.~G.,  {Marchesini} D.,
  {Labb{\'e}} I.,  {Franx} M.,  {Kriek} M.,  {Quadri} R.~F.,  {Illingworth} G.,
   {Lee} K.,  {Muzzin} A.,    {Rudnick} G.,  2009, \apjl, 706, L173

\bibitem[\protect\citeauthoryear{{Bruzual} \& {Charlot}}{{Bruzual} \&
  {Charlot}}{2003}]{BC03}
{Bruzual} G.,  {Charlot} S.,  2003, \mnras, 344, 1000

\bibitem[\protect\citeauthoryear{{Doherty}, {Tanaka}, {De Breuck}, {Ly},
  {Kodama}, {Kurk}, {Seymour}, {Vernet}, {Stern}, {Venemans}, {Kajisawa} \&
  {Tanaka}}{{Doherty} et~al.}{2010}]{Doherty2010}
{Doherty} M.,  {Tanaka} M.,  {De Breuck} C.,  {Ly} C.,  {Kodama} T.,  {Kurk}
  J.,  {Seymour} N.,  {Vernet} J.,  {Stern} D.,  {Venemans} B.,  {Kajisawa} M.,
     {Tanaka} I.,  2010, \aap, 509, A83+

\bibitem[\protect\citeauthoryear{{Dressler}}{{Dressler}}{1980}]{Dressler1980}
{Dressler} A.,  1980, \apj, 236, 351

\bibitem[\protect\citeauthoryear{{Ellis}, {Smail}, {Dressler}, {Couch},
  {Oemler} Jr., {Butcher} \& {Sharples}}{{Ellis} et~al.}{1997}]{Ellis1997}
{Ellis} R.~S.,  {Smail} I.,  {Dressler} A.,  {Couch} W.~J.,  {Oemler} Jr. A.,
  {Butcher} H.,    {Sharples} R.~M.,  1997, \apj, 483, 582

\bibitem[\protect\citeauthoryear{{Finger}, {Dorn}, {Eschbaumer}, {Hall},
  {Mehrgan}, {Meyer} \& {Stegmeier}}{{Finger} et~al.}{2008}]{finger2008}
{Finger} G.,  {Dorn} R.~J.,  {Eschbaumer} S.,  {Hall} D.~N.~B.,  {Mehrgan} L.,
  {Meyer} M.,    {Stegmeier} J.,  2008, in Society of Photo-Optical
  Instrumentation Engineers (SPIE) Conference Series Vol.~7021 of Presented at
  the Society of Photo-Optical Instrumentation Engineers (SPIE) Conference,
  {Performance evaluation, readout modes, and calibration techniques of HgCdTe
  Hawaii-2RG mosaic arrays}

\bibitem[\protect\citeauthoryear{{Galametz}, {Vernet}, {De Breuck}, {Hatch},
  {Miley}, {Kodama}, {Kurk}, {Overzier}, {Rettura}, {Rottgering}, {Seymour},
  {Venemans} \& {Zirm}}{{Galametz} et~al.}{2010}]{Galametz2010}
{Galametz} A.,  {Vernet} J.,  {De Breuck} C.,  {Hatch} N.,  {Miley} G.,
  {Kodama} T.,  {Kurk} J.,  {Overzier} R.,  {Rettura} A.,  {Rottgering} H.,
  {Seymour} N.,  {Venemans} B.,    {Zirm} A.,  2010, ArXiv e-prints

\bibitem[\protect\citeauthoryear{{Gladders}, {Lopez-Cruz}, {Yee} \&
  {Kodama}}{{Gladders} et~al.}{1998}]{Gladders1998}
{Gladders} M.~D.,  {Lopez-Cruz} O.,  {Yee} H.~K.~C.,    {Kodama} T.,  1998,
  \apj, 501, 571

\bibitem[\protect\citeauthoryear{{Gunn} \& {Gott} III}{{Gunn} \&
  {Gott}}{1972}]{Gunn1972}
{Gunn} J.~E.,  {Gott} III J.~R.,  1972, \apj, 176, 1

\bibitem[\protect\citeauthoryear{{Infante}}{{Infante}}{1994}]{Infante1994}
{Infante} L.,  1994, \aap, 282, 353

\bibitem[\protect\citeauthoryear{{Kajisawa}, {Kodama}, {Tanaka}, {Yamada} \&
  {Bower}}{{Kajisawa} et~al.}{2006}]{Kajisawa2006}
{Kajisawa} M.,  {Kodama} T.,  {Tanaka} I.,  {Yamada} T.,    {Bower} R.,  2006,
  \mnras, 371, 577

\bibitem[\protect\citeauthoryear{{Kissler-Patig}, {Pirard}, {Casali},
  {Moorwood}, {Ageorges}, {Alves de Oliveira}, {Baksai}, {Bedin}, {Bendek},
  {Biereichel}, {Delabre}, {Dorn}, {Esteves} \& {et al.}}{{Kissler-Patig}
  et~al.}{2008}]{Kissler-Patig2008}
{Kissler-Patig} M.,  {Pirard} J.,  {Casali} M.,  {Moorwood} A.,  {Ageorges} N.,
   {Alves de Oliveira} C.,  {Baksai} P.,  {Bedin} L.~R.,  {Bendek} E.,
  {Biereichel} P.,  {Delabre} B.,  {Dorn} R.,  {Esteves} R.,    {et al.} 2008,
  \aap, 491, 941

\bibitem[\protect\citeauthoryear{{Kodama}, {Tanaka}, {Kajisawa}, {Kurk},
  {Venemans}, {De Breuck}, {Vernet} \& {Lidman}}{{Kodama}
  et~al.}{2007}]{Kodama2007}
{Kodama} T.,  {Tanaka} I.,  {Kajisawa} M.,  {Kurk} J.,  {Venemans} B.,  {De
  Breuck} C.,  {Vernet} J.,    {Lidman} C.,  2007, \mnras, 377, 1717

\bibitem[\protect\citeauthoryear{{Kurk}, {Cimatti}, {Zamorani}, {Halliday},
  {Mignoli}, {Pozzetti}, {Daddi}, {Rosati}, {Dickinson}, {Bolzonella},
  {Cassata}, {Renzini}, {Franceschini}, {Rodighiero} \& {Berta}}{{Kurk}
  et~al.}{2009}]{Kurk2009}
{Kurk} J.,  {Cimatti} A.,  {Zamorani} G.,  {Halliday} C.,  {Mignoli} M.,
  {Pozzetti} L.,  {Daddi} E.,  {Rosati} P.,  {Dickinson} M.,  {Bolzonella} M.,
  {Cassata} P.,  {Renzini} A.,  {Franceschini} A.,  {Rodighiero} G.,    {Berta}
  S.,  2009, \aap, 504, 331

\bibitem[\protect\citeauthoryear{{Kurk}, {Pentericci}, {R{\"o}ttgering} \&
  {Miley}}{{Kurk} et~al.}{2004}]{Kurk2004}
{Kurk} J.~D.,  {Pentericci} L.,  {R{\"o}ttgering} H.~J.~A.,    {Miley} G.~K.,
  2004, \aap, 428, 793

\bibitem[\protect\citeauthoryear{{Kurk}, {R{\"o}ttgering}, {Pentericci},
  {Miley}, {van Breugel}, {Carilli}, {Ford}, {Heckman}, {McCarthy} \&
  {Moorwood}}{{Kurk} et~al.}{2000}]{Kurk2000}
{Kurk} J.~D.,  {R{\"o}ttgering} H.~J.~A.,  {Pentericci} L.,  {Miley} G.~K.,
  {van Breugel} W.,  {Carilli} C.~L.,  {Ford} H.,  {Heckman} T.,  {McCarthy}
  P.,    {Moorwood} A.,  2000, \aap, 358, L1

\bibitem[\protect\citeauthoryear{{Landy} \& {Szalay}}{{Landy} \&
  {Szalay}}{1993}]{Landy1993}
{Landy} S.~D.,  {Szalay} A.~S.,  1993, \apj, 412, 64

\bibitem[\protect\citeauthoryear{{Lawrence}}{{Lawrence}}{2007}]{Lawrence2007}
{Lawrence} A. e.~a.,  2007, \mnras, 379, 1599

\bibitem[\protect\citeauthoryear{{Le Fevre}, {Deltorn}, {Crampton} \&
  {Dickinson}}{{Le Fevre} et~al.}{1996}]{LeFevre1996}
{Le Fevre} O.,  {Deltorn} J.~M.,  {Crampton} D.,    {Dickinson} M.,  1996,
  \apjl, 471, L11+

\bibitem[\protect\citeauthoryear{{Ling}, {Barrow} \& {Frenk}}{{Ling}
  et~al.}{1986}]{Ling1986}
{Ling} E.~N.,  {Barrow} J.~D.,    {Frenk} C.~S.,  1986, \mnras, 223, 21P

\bibitem[\protect\citeauthoryear{{Mei}, {Holden}, {Blakeslee}, {Ford}, {Franx},
  {Homeier}, {Illingworth}, {Jee}, {Overzier}, {Postman}, {Rosati}, {Van der
  Wel} \& {Bartlett}}{{Mei} et~al.}{2009}]{Mei2009}
{Mei} S.,  {Holden} B.~P.,  {Blakeslee} J.~P.,  {Ford} H.~C.,  {Franx} M.,
  {Homeier} N.~L.,  {Illingworth} G.~D.,  {Jee} M.~J.,  {Overzier} R.,
  {Postman} M.,  {Rosati} P.,  {Van der Wel} A.,    {Bartlett} J.~G.,  2009,
  \apj, 690, 42

\bibitem[\protect\citeauthoryear{{Miley} \& {De Breuck}}{{Miley} \& {De
  Breuck}}{2008}]{MileydeBreuck2008}
{Miley} G.,  {De Breuck} C.,  2008, \araa, pp~1--+

\bibitem[\protect\citeauthoryear{{Monet}, {Levine}, {Canzian}, {Ables}, {Bird},
  {Dahn}, {Guetter}, {Harris}, {Henden}, {Leggett}, {Levison}, {Luginbuhl},
  {Martini}, {Monet}, {Munn}, {Pier}, {Rhodes} \& {et al.}}{{Monet}
  et~al.}{2003}]{Monet2003}
{Monet} D.~G.,  {Levine} S.~E.,  {Canzian} B.,  {Ables} H.~D.,  {Bird} A.~R.,
  {Dahn} C.~C.,  {Guetter} H.~H.,  {Harris} H.~C.,  {Henden} A.~A.,  {Leggett}
  S.~K.,  {Levison} H.~F.,  {Luginbuhl} C.~B.,  {Martini} J.,  {Monet}
  A.~K.~B.,  {Munn} J.~A.,  {Pier} J.~R.,  {Rhodes} A.~R.,    {et al.} 2003,
  \aj, 125, 984

\bibitem[\protect\citeauthoryear{{Overzier}, {Bouwens}, {Cross}, {Venemans},
  {Miley}, {Zirm}, {Ben{\'{\i}}tez}, {Blakeslee}, {Coe}, {Demarco}, {Ford} \&
  {Homeier}}{{Overzier} et~al.}{2008}]{Overzier2008}
{Overzier} R.~A.,  {Bouwens} R.~J.,  {Cross} N.~J.~G.,  {Venemans} B.~P.,
  {Miley} G.~K.,  {Zirm} A.~W.,  {Ben{\'{\i}}tez} N.,  {Blakeslee} J.~P.,
  {Coe} D.,  {Demarco} R.,  {Ford} H.~C.,    {Homeier} N.~L.,  2008, \apj, 673,
  143

\bibitem[\protect\citeauthoryear{{Overzier}, {Miley}, {Bouwens}, {Cross},
  {Zirm}, {Ben{\'{\i}}tez}, {Blakeslee} \& {Clampin}}{{Overzier}
  et~al.}{2006}]{Overzier2006}
{Overzier} R.~A.,  {Miley} G.~K.,  {Bouwens} R.~J.,  {Cross} N.~J.~G.,  {Zirm}
  A.~W.,  {Ben{\'{\i}}tez} N.,  {Blakeslee} J.~P.,    {Clampin} M. e.~a.,
  2006, \apj, 637, 58

\bibitem[\protect\citeauthoryear{{Papovich}, {Momcheva}, {Willmer},
  {Finkelstein}, {Finkelstein}, {Tran}, {Brodwin}, {Dunlop}, {Farrah}, {Khan},
  {Lotz}, {McCarthy}, {McLure}, {Rieke}, {Rudnick} \& {Sivanandam}}{{Papovich}
  et~al.}{2010}]{Papovich2010}
{Papovich} C.,  {Momcheva} I.,  {Willmer} C.~N.~A.,  {Finkelstein} K.~D.,
  {Finkelstein} S.~L.,  {Tran} K.,  {Brodwin} M.,  {Dunlop} J.~S.,  {Farrah}
  D.,  {Khan} S.~A.,  {Lotz} J.,  {McCarthy} P.,  {McLure} R.~J.,  {Rieke} M.,
  {Rudnick} G.,    {Sivanandam} S.,  2010, ArXiv e-prints

\bibitem[\protect\citeauthoryear{{Peacock}}{{Peacock}}{1999}]{Peacock1999}
{Peacock} J.~A.,  1999, {Cosmological Physics}.
Cambridge University Press

\bibitem[\protect\citeauthoryear{{Pentericci}, {Kurk}, {R{\"o}ttgering},
  {Miley}, {van Breugel}, {Carilli}, {Ford}, {Heckman}, {McCarthy} \&
  {Moorwood}}{{Pentericci} et~al.}{2000}]{Pentericci2000}
{Pentericci} L.,  {Kurk} J.~D.,  {R{\"o}ttgering} H.~J.~A.,  {Miley} G.~K.,
  {van Breugel} W.,  {Carilli} C.~L.,  {Ford} H.,  {Heckman} T.,  {McCarthy}
  P.,    {Moorwood} A.,  2000, \aap, 361, L25

\bibitem[\protect\citeauthoryear{{Quadri}, {van Dokkum}, {Gawiser}, {Franx},
  {Marchesini}, {Lira}, {Rudnick}, {Herrera}, {Maza}, {Kriek}, {Labb{\'e}} \&
  {Francke}}{{Quadri} et~al.}{2007}]{Quadri2007}
{Quadri} R.,  {van Dokkum} P.,  {Gawiser} E.,  {Franx} M.,  {Marchesini} D.,
  {Lira} P.,  {Rudnick} G.,  {Herrera} D.,  {Maza} J.,  {Kriek} M.,
  {Labb{\'e}} I.,    {Francke} H.,  2007, \apj, 654, 138

\bibitem[\protect\citeauthoryear{{Rocca-Volmerange}, {Le Borgne}, {De Breuck},
  {Fioc} \& {Moy}}{{Rocca-Volmerange} et~al.}{2004}]{Rocca-Volmerange2004}
{Rocca-Volmerange} B.,  {Le Borgne} D.,  {De Breuck} C.,  {Fioc} M.,    {Moy}
  E.,  2004, \aap, 415, 931

\bibitem[\protect\citeauthoryear{{Roche}, {Eales}, {Hippelein} \&
  {Willott}}{{Roche} et~al.}{1999}]{Roche1999}
{Roche} N.,  {Eales} S.~A.,  {Hippelein} H.,    {Willott} C.~J.,  1999, \mnras,
  306, 538

\bibitem[\protect\citeauthoryear{{Seymour}, {Stern}, {De Breuck}, {Vernet},
  {Rettura}, {Dickinson}, {Dey}, {Eisenhardt}, {Fosbury}, {Lacy}, {McCarthy},
  {Miley}, {Rocca-Volmerange}, {Rottgering}, {Stanford}, {Teplitz} \& {van
  Breugel}}{{Seymour} et~al.}{2007}]{Seymour2007}
{Seymour} N.,  {Stern} D.,  {De Breuck} C.,  {Vernet} J.,  {Rettura} A.,
  {Dickinson} M.,  {Dey} A.,  {Eisenhardt} P.,  {Fosbury} R.,  {Lacy} M.,
  {McCarthy} P.,  {Miley} G.,  {Rocca-Volmerange} B.,  {Rottgering} H.,
  {Stanford} S.~A.,  {Teplitz} H.,    {van Breugel} W.,  2007, ArXiv
  Astrophysics e-prints

\bibitem[\protect\citeauthoryear{{Stanford}, {Eisenhardt} \&
  {Dickinson}}{{Stanford} et~al.}{1998}]{Stanford1998}
{Stanford} S.~A.,  {Eisenhardt} P.~R.,    {Dickinson} M.,  1998, \apj, 492, 461

\bibitem[\protect\citeauthoryear{{Steidel}, {Adelberger}, {Dickinson},
  {Giavalisco}, {Pettini} \& {Kellogg}}{{Steidel} et~al.}{1998}]{Steidel1998}
{Steidel} C.~C.,  {Adelberger} K.~L.,  {Dickinson} M.,  {Giavalisco} M.,
  {Pettini} M.,    {Kellogg} M.,  1998, \apj, 492, 428

\bibitem[\protect\citeauthoryear{{Steidel}, {Adelberger}, {Shapley}, {Erb},
  {Reddy} \& {Pettini}}{{Steidel} et~al.}{2005}]{Steidel2005}
{Steidel} C.~C.,  {Adelberger} K.~L.,  {Shapley} A.~E.,  {Erb} D.~K.,  {Reddy}
  N.~A.,    {Pettini} M.,  2005, \apj, 626, 44

\bibitem[\protect\citeauthoryear{{Tanaka}, {Finoguenov} \& {Ueda}}{{Tanaka}
  et~al.}{2010}]{Tanaka2010}
{Tanaka} M.,  {Finoguenov} A.,    {Ueda} Y.,  2010, ArXiv e-prints

\bibitem[\protect\citeauthoryear{{van Dokkum}, {Quadri}, {Marchesini},
  {Rudnick}, {Franx}, {Gawiser}, {Herrera}, {Wuyts}, {Lira}, {Labb{\'e}} \&
  {Maza}}{{van Dokkum} et~al.}{2006}]{vanDokkum2006}
{van Dokkum} P.~G.,  {Quadri} R.,  {Marchesini} D.,  {Rudnick} G.,  {Franx} M.,
   {Gawiser} E.,  {Herrera} D.,  {Wuyts} S.,  {Lira} P.,  {Labb{\'e}} I.,
  {Maza} J.,  2006, \apjl, 638, L59

\bibitem[\protect\citeauthoryear{{van Dokkum} \& {van der Marel}}{{van Dokkum}
  \& {van der Marel}}{2007}]{vanDokkum2007}
{van Dokkum} P.~G.,  {van der Marel} R.~P.,  2007, \apj, 655, 30

\bibitem[\protect\citeauthoryear{Vandame}{Vandame}{2004}]{Vandame2004}
Vandame B.,  2004, PhD Thesis, Nice University, France

\bibitem[\protect\citeauthoryear{{Venemans}, {R{\"o}ttgering}, {Miley}, {van
  Breugel}, {de Breuck}, {Kurk}, {Pentericci}, {Stanford}, {Overzier}, {Croft}
  \& {Ford}}{{Venemans} et~al.}{2007}]{Venemans2007}
{Venemans} B.~P.,  {R{\"o}ttgering} H.~J.~A.,  {Miley} G.~K.,  {van Breugel}
  W.~J.~M.,  {de Breuck} C.,  {Kurk} J.~D.,  {Pentericci} L.,  {Stanford}
  S.~A.,  {Overzier} R.~A.,  {Croft} S.,    {Ford} H.,  2007, \aap, 461, 823

\bibitem[\protect\citeauthoryear{{Visvanathan} \& {Sandage}}{{Visvanathan} \&
  {Sandage}}{1977}]{Visvanathan1977}
{Visvanathan} N.,  {Sandage} A.,  1977, \apj, 216, 214

\bibitem[\protect\citeauthoryear{{Williams}, {Quadri}, {Franx}, {van Dokkum} \&
  {Labb{\'e}}}{{Williams} et~al.}{2009a}]{Williams2009a}
{Williams} R.~J.,  {Quadri} R.~F.,  {Franx} M.,  {van Dokkum} P.,
  {Labb{\'e}} I.,  2009, \apj, 691, 1879

\bibitem[\protect\citeauthoryear{{Williams}, {Quadri}, {Franx}, {van Dokkum},
  {Toft}, {Kriek} \& {Labbe}}{{Williams} et~al.}{2009b}]{Williams2009b}
{Williams} R.~J.,  {Quadri} R.~F.,  {Franx} M.,  {van Dokkum} P.,  {Toft} S.,
  {Kriek} M.,    {Labbe} I.,  2009, ArXiv e-prints

\bibitem[\protect\citeauthoryear{{Wuyts}, {Franx}, {Cox}, {F{\"o}rster
  Schreiber}, {Hayward}, {Hernquist}, {Hopkins}, {Labb{\'e}}, {Marchesini},
  {Robertson}, {Toft} \& {van Dokkum}}{{Wuyts} et~al.}{2009}]{Wuyts2009}
{Wuyts} S.,  {Franx} M.,  {Cox} T.~J.,  {F{\"o}rster Schreiber} N.~M.,
  {Hayward} C.~C.,  {Hernquist} L.,  {Hopkins} P.~F.,  {Labb{\'e}} I.,
  {Marchesini} D.,  {Robertson} B.~E.,  {Toft} S.,    {van Dokkum} P.~G.,
  2009, \apj, 700, 799

\bibitem[\protect\citeauthoryear{{Wuyts}, {Labb{\'e}}, {Franx}, {Rudnick} \&
  {van Dokkum}}{{Wuyts} et~al.}{2007}]{Wuyts2007}
{Wuyts} S.,  {Labb{\'e}} I.,  {Franx} M.,  {Rudnick} G.,    {van Dokkum} e.~a.,
   2007, \apj, 655, 51

\end{thebibliography}
\label{lastpage}
\clearpage
\end{document}